\documentclass[preprint,showpacs,preprintnumbers,amsmath,amssymb,showkeys]{revtex4}

\usepackage{graphicx}
\usepackage{dcolumn}
\usepackage{bm}
\usepackage{braket}
  
\begin{document} 

\noindent

\preprint{}

\title{Quantum uncertainty as classical uncertainty of real-deterministic variables constructed from complex weak values and a global random variable}
 
\author{Agung Budiyono$^{1,2,3,4}$} 
\email{agungbymlati@gmail.com}
\author{Hermawan K. Dipojono$^{1,2}$}
\affiliation{$^1$Research Center for Nanoscience and Nanotechnology, Bandung Institute of Technology, Bandung, 40132 Indonesia}
\affiliation{$^2$Department of Engineering Physics, Bandung Institute of Technology, Bandung, 40132 Indonesia}
\affiliation{$^3$Laboratory for Quantum Technology, School of Electrical Engineering and Informatics, Bandung Institute of Technology, Bandung, 40132, Indonesia}
\affiliation{$^4$Kubus Computing and Research, Juwana, Pati, 59185 Indonesia}
 
\date{\today}
  
\begin{abstract}   

What does it take for real-deterministic c-valued (i.e., classical, commuting) variables to comply with the Heisenberg uncertainty principle? Here, we construct a class of real-deterministic c-valued variables out of the weak values obtained via a non-perturbing weak measurement of quantum operators with a post-selection over a complete set of state vectors basis, which always satisfies the Kennard-Robertson-Schr\"odinger uncertainty relation. First, we introduce an auxiliary global random variable and couple it to the imaginary part of the weak value to transform the incompatibility between the quantum operator and the basis into the fluctuation of an `error term', and then superimpose it onto the real-part of the weak value. We show that this class of ``c-valued physical quantities'' provides a real-deterministic contextual hidden variable model for the quantum expectation value of a certain class of operators. We then show that the Schr\"odinger and the Kennard-Robertson lower bounds can be obtained separately by imposing the classical uncertainty relation to the c-valued physical quantities associated with a pair of Hermitian operators. Within the representation, the complementarity between two incompatible quantum observables manifests the absence of a basis wherein the error terms of the associated two c-valued physical quantities simultaneously vanish. Furthermore, quantum uncertainty relation is captured by a specific irreducible epistemic restriction, foreign in classical mechanics, constraining the allowed form of the joint distribution of the two c-valued physical quantities. We then suggest an epistemic interpretation of the two terms decomposing the c-valued physical quantity as the optimal estimate under the epistemic restriction and the associated estimation error, and discuss the classical limit.  
  
\end{abstract}    

\maketitle                                       

\section{Introduction}

Among the many profound and bizarre implications of the fundamental noncommutativity between the observables in quantum mechanics, the Heisenberg uncertainty principle stands to offer the most direct physical intuition about the quantum randomness  \cite{Heisenberg UR}. The Kennard-Robertson-Schr\"odinger (KRS) uncertainty relation \cite{Kennard UR,Robertson UR,Schroedinger UR} | one of the manifestations of the quantum uncertainty principle | dictates that the product of the standard deviations of two observables is fundamentally bounded from below by a value determined by the commutation relation between the two observables. Since the beginning of quantum mechanics, some researchers have speculated that quantum randomness might be explained by deterministic variables associated with noncommuting quantum observables. The most famous argument in favour of such an approach is given by Einstein, Podolsky and Rosen in their seminal paper \cite{EPR paradox}. Bell \cite{Bell's theorem,Bell contextuality paper} and Kochen and Specker \cite{KS contextuality paper} later showed, however, that, to comply with certain specific quantum predictions, such deterministic variables must violate some of the classically plausible assumptions about Nature. Here we ask: what classical assumptions we need to give in to construct real-valued deterministic variables associated with noncommuting quantum observables so that they respect the KRS uncertainty relation. One way to study this problem is to construct such real-valued deterministic variables, and to identify the fundamental relation among them | foreign in classical mechanics | which capture the KRS uncertainty relation. Such a relation, if exists, cannot be seen directly via the standard strong (i.e., sharp) measurement in quantum mechanics since it does not allow a joint measurement of two noncommuting observables: a sharp measurement of one observable in general disturbs the system, making the observable noncommuting with the first becomes indefinite (indeterministic). 

On the other hand, a novel scheme of measurement was introduced three decades ago by Aharonov, Albert and Vaidman which goes beyond the measurement-induced-disturbance in standard measurement \cite{Aharonov weak value,Ahoronov-Rohrlich book}. In this scheme, given a pre-selected ensemble characterized by a quantum state $\ket{\psi}$, one first weakly measures a quantum observable $\hat{O}$ without significantly perturbing the quantum state. This is then followed with a post-selection over an ensemble characterized by $\ket{\phi}$ via a strong measurement which may be incompatible with the first weak measurement. Repeating the process over the pre- and post-selected ensembles, one gets a weak value: $O^{\rm w}(\phi|\psi)\doteq\frac{\braket{\phi|\hat{O}|\psi}}{\braket{\phi|\psi}}$, where it is assumed that $\braket{\phi|\psi}\neq 0$. Using the measurement procedure, one may thus, to an extent, obtain simultaneous information about the two incompatible observables, otherwise inaccessible via the standard strong measurement. The real and imaginary parts of the weak value can be inferred respectively from the average shift of the position and momentum of the pointer of the measuring device \cite{Lundeen complex weak value,Jozsa complex weak value}. The observation of the weak value via an arbitrarily nonperturbing weak measurement with post-selection can also be argued to reflect our ordinary and naive idea of measurement in the presence of randomness \cite{Wiseman Bohmian mechanics}. Moreover, the fact that it is obtained via weak measurement intuitively suggests that it somehow reveals the counterfactual properties of the system compatible with pre-selection and post-selection \cite{Wiseman Bohmian mechanics,Aharonov weak value as objective counterfactual value 1,Aharonov weak value as objective counterfactual value 2,Tollaksen weak value as objective counterfactual value,Hosoya-Koga weak value as objective counterfactual value,Hosoya-Shikano counterfactual value,Steinberg tunneling time via weak value 1,Steinberg tunneling time via weak value 2,Hofmann more complete description}. This insight has motivated the application of the concept of weak value to study various quantum conundrum \cite{Aharonov weak value Hardy paradox,Lundeen weak value Hardy paradox,Hosoya-Shikano counterfactual value,Yokota weak value Hardy paradox,Molmer negative population,Resch quantum box problem,Aharonov QCC,Matzkin QCC,Denkmayr QCC,Wiseman weak valued distribution momentum transfer,Steinberg average trajectory,Williams Leggett-Garg inequality,Steinberg tunneling time via weak value 1,Steinberg tunneling time via weak value 2}, and the meaning of a quantum state \cite{Lundeen direct measurement of the quantum state,Salvail direct measurement of quantum state,Agung epistemic interpretation}. We note that the real or/and imaginary parts of the weak value can also naturally arise in schemes not involving weak measurements \cite{Hall exact UR,Hall prior information,Agung epistemic interpretation,Johansen weak value best estimation,Hofmann imaginary part of weak value in optimal estimation,Johansen reconstruction of weak value from strong measurement,Bliokh weak value in classical wave optics,Holland local expectation value}, suggesting that it might have a deep nature beyond the operational context of weak measurement with post-selection. With all this in mind, it is interesting to ask to what extent the information contained in weak values is sufficient to construct deterministic variables associated with the noncommuting observables which capture the KRS uncertainty relation. 

Indeed, some significant efforts have been made to reconstruct quantum statistics based on deterministic variables by using the weak values. It has been argued, for example, that weak values can capture the contextuality of the quantum fluctuations \cite{Tollaksen contextual value - weak value,Dressel contextual value - weak value,Hofmann contextual value - weak value}. In particular, in Ref. \cite{Hofmann contextual value - weak value}, it is shown that the complex weak values explicitly reveal the contextuality of quantum fluctuations that are defined by the choice of the measurement post-selection and the quantum state preparation. Moreover, such deterministic weak values can also be used to derive the KRS uncertainty relation by subjecting them to classical uncertainty relation \cite{Hall weak value to quantum uncertainty}. Note however that these weak values are in general complex even if the associated quantum operators are Hermitian. Such complex deterministic variables do not give a transparent insight into the fundamental relation among them which underlies the KRS uncertainty relation. Furthermore, it is desirable to have real-valued deterministic variables which not only smoothly generalize the strong values given by the real eigenvalues of Hermitian observables, but which also offer insight into a conceptually smooth quantum-classical correspondence. 

In the present work, we discuss the above problems by devising a class of deterministic real-valued variables constructed from complex weak values which always satisfy the KRS uncertainty relation. This work essentially extends the results reported in Refs. \cite{Agung-Daniel model,Agung ERPS representation} to finite dimensional discrete variable quantum systems. First, in Sec. \ref{c-valued physical quantities from weak values and an auxiliary global random variable}, given a quantum operator, a quantum state associated with a preparation, and a complete set of state vectors in the Hilbert space as a reference coordinate basis, we construct a real-deterministic c-valued (i.e., classical, commuting) variable from the associated complex weak value with a post-selection over the basis, which we call `c-valued physical quantity'. See Eq. (\ref{fundamental epistemic decomposition}). To do this, we introduce an auxiliary real-valued global-nonseparable (i.e., spatially uniform) random variable $\xi$ and couple it to the imaginary part of the weak value, transforming the incompatibility between the quantum operator and the reference coordinate basis into the strength of the fluctuations of $\xi$. This $\xi$-dependent `error term' is then superimposed onto a $\xi$-independent term given by the real part of the weak value. We show in Sec. \ref{c-valued physical quantities as deterministic hidden variable models for quantum expectation values of products operators} that this class of c-valued physical quantities provide a real-deterministic contextual hidden variable model for the quantum expectation value of a class of quantum operators. The contexts underlying the c-valued physical quantities are given by the choice of the reference basis which defines the post-selection in the weak value measurement, the choice of weak measurement when the associated observable can be written into a product of two observables, and the choice of preparation associated with a given density matrix (mixed state).  

In Sec. \ref{Quantum uncertainty relation from real deterministic c-valued physical quantities}, we show that varying the reference basis, the quantum variance of a Hermitian operator over a quantum state is decomposed into the trade-off between the strength of the $\xi$-independent and $\xi$-dependent fluctuations of the associated c-valued physical quantity. The Schr\"odinger and the Kennard-Robertson lower bounds in quantum uncertainty relation can then be obtained separately by imposing the classical uncertainty relation respectively to the $\xi$-independent term and the $\xi$-dependent error term. In this representation, the finite Kennard-Robertson lower bound associated with two incompatible quantum observables reflects the absence of a common reference coordinate basis wherein the $\xi-$dependent error terms of the associated two c-valued physical quantities simultaneously vanish. We proceed in Sec. \ref{Quantum incompatibility as epistemic restriction} to argue that the formalism captures the incompatibility between two observables in terms of a specific fundamental epistemic restriction between the associated two c-valued physical quantities. It constrains the allowed form of their joint distribution and thereby underlies the KRS uncertainty relation. In this sense, a specific form of epistemic freedom, fundamental in classical mechanics, has to be given in. We then suggest in Sec. \ref{Optimal estimation under epistemic restriction} an epistemic interpretation of the two terms decomposing the c-valued physical quantity as the optimal estimate under epistemic restriction and the associated single-shot estimation error, and discuss the classical limit. Sec. \ref{conclusion} summarizes the present paper and offers a future outlook. 

\section{Real-deterministic c-valued physical quantities}

\subsection{Real-deterministic physical quantities from complex weak values and an auxiliary global random variable \label{c-valued physical quantities from weak values and an auxiliary global random variable}}  

First, given the Hilbert space associated with the quantum system under interest, let us choose a complete set of quantum states $\{\ket{\phi_n}\}$ satisfying the completeness relation $\sum_n\ket{\phi_n}\bra{\phi_n}=\hat{1}$, where the summation is replaced with a suitable integration for a continuous set, and $\hat{1}$ is the identity matrix. We take this complete set of vectors in the Hilbert space as the reference coordinate basis. We then define a vector of weak values $O^{\rm w}(\phi_n|\psi)$ associated with an operator $\hat{O}$ with a pre-selected state $\ket{\psi}$ and a post-selection over the complete set of quantum states $\{\ket{\phi_n}\}$ as
\begin{eqnarray}
O^{\rm w}(\phi_n|\psi)\doteq\frac{\braket{\phi_n|\hat{O}|\psi}}{\braket{\phi_n|\psi}}. 
\label{complex weak value: definition}
\end{eqnarray}
Let us denote the real and imaginary parts of $O^{\rm w}(\phi_n|\psi)$ respectively as
\begin{eqnarray}
O^{\rm w}_{\mathcal{R}}(\phi_n|\psi)&\doteq&{\rm Re}\big\{O^{\rm w}(\phi_n|\psi)\big\}=\frac{\braket{\psi|\{\hat{\Pi}_{\phi_n},\hat{O}\}|\psi}}{2|\braket{\phi_n|\psi}|^2},\nonumber\\
O^{\rm w}_{\mathcal{I}}(\phi_n|\psi)&\doteq&{\rm Im}\big\{O^{\rm w}(\phi_n|\psi)\big\}=\frac{\braket{\psi|[\hat{\Pi}_{\phi_n},\hat{O}]|\psi}}{2i|\braket{\phi_n|\psi}|^2},
\label{real and imaginary parts of weak value vs compatibility and incompatibility}
\end{eqnarray}
where $\hat{\Pi}_{\phi}\doteq\ket{\phi}\bra{\phi}$, $\{\hat{A},\hat{B}\}\doteq \hat{A}\hat{B}+\hat{B}^{\dagger}\hat{A}^{\dagger}$ and $[\hat{A},\hat{B}]\doteq\hat{A}\hat{B}-\hat{B}^{\dagger}\hat{A}^{\dagger}$. Notice in particular that when $\hat{O}$ in Eq. (\ref{real and imaginary parts of weak value vs compatibility and incompatibility}) is Hermitian, i.e., $\hat{O}=\hat{O}^{\dagger}$, the imaginary part of the weak value incorporates the incompatibility between the quantum observable $\hat{O}$ and the projectors  $\{\Pi_{\phi_n}\}$ associated with the reference coordinate basis $\{\ket{\phi_n}\}$ via their commutation relation, while the real part takes the form of an anticommutation relation. By definition, unlike the standard strong value obtained in standard strong measurement (without post-selection) which is in general indefinite with respect to the measurement basis $\{\ket{\phi_n}\}$, the weak value is thus completely determined by the choice of the coordinate basis, i.e., $\phi_n\mapsto O^{\rm w}(\phi_n|\psi)$. It should be noted importantly that, while the strong value refers to a single event (i.e., a single repetition of a strong measurement), the weak value is obtained by repeating the weak measurement over the pre- and post-selected ensembles and taking an average. 

Next, for each quantum operator $\hat{O}$ and a preparation characterized by a quantum state $\ket{\psi}$, let us define a real-deterministic c-valued physical quantity $\tilde{O}(\phi_n,\xi|\psi)$ which depends on the choice of a reference coordinate basis $\{\ket{\phi_n}\}$ as 
\begin{eqnarray}
\tilde{O}(\phi_n,\xi|\psi)\doteq O^{\rm w}_{\mathcal{R}}(\phi_n|\psi)+\frac{\xi}{\hbar}O^{\rm w}_{\mathcal{I}}(\phi_n|\psi). 
\label{fundamental epistemic decomposition}
\end{eqnarray} 
Here, $\xi$ is a global-nonseparable variable assumed to fluctuate randomly on a microscopic time scale independent of the pre-selected state $\ket{\psi}$, the operator $\hat{O}$, and the reference coordinate basis $\{\ket{\phi_n}\}$. For example, we can assume that each single run of the weak measurement is assigned a single random realization of $\xi$. By global-nonseparable, we meant $\xi$ is spatially uniform so that, two systems, arbitrarily spatially  separated from each other, always experience the same simultaneous value of $\xi$. Moreover, we assume that its average and variance are respectively given by \cite{Agung-Daniel model} 
\begin{equation}
\overline{\xi}\doteq\int{\rm d}\xi\xi\chi(\xi)=0,~~\overline{\xi^2}=\hbar^2, 
\label{Planck constant}
\end{equation}
where $\chi(\xi)$ is the probability density of $\xi$. We shall here on refer to the $\xi$-dependent term on the right-hand side of Eq. (\ref{fundamental epistemic decomposition}) as the `error term', suggesting that it is an epistemic rather than an ontic quantity. The reason for this naming will be made clear in Sec. \ref{Optimal estimation under epistemic restriction}. An extension of the above defined c-valued physical quantity to a pre-selected mixed quantum state (density matrix) is given at the end of Sec. \ref{c-valued physical quantities as deterministic hidden variable models for quantum expectation values of products operators}. 

Operationally, $\tilde{O}(\phi_n,\xi|\psi)$ defined in Eq. (\ref{fundamental epistemic decomposition}) can thus be obtained as follows. First, one makes weak measurement of $\hat{O}$ with a pre-selected state $\ket{\psi}$ and a post-selected state $\ket{\phi_n}$ and infers the real and imaginary parts of the weak value from the average position and momentum of the pointer of the measuring device. Next, one generates $\xi$ using a random number generator, i.e., sampling $\xi$ from $\chi(\xi)$, satisfying Eq. (\ref{Planck constant}), e.g., $\xi=\pm\hbar$ with half-half probability. Given the pair of $(\phi_n,\xi)$, one then computes $\tilde{O}(\phi_n,\xi|\psi)$ as prescribed in Eq. (\ref{fundamental epistemic decomposition}). At present, the c-valued physical quantities defined above thus cannot be observed directly via weak measurement but involves classical data processing, similar to the other meaningful physical quantities like quantum entanglement. Conversely, from Eq. (\ref{fundamental epistemic decomposition}), given $\tilde{O}(\phi_n,\xi|\psi)$, the real and imaginary parts of the weak value $O^{\rm w}(\phi_n|\psi)$ can be recovered as $O^{\rm w}_{\mathcal{R}}(\phi_n|\psi)=\frac{1}{2}(\tilde{O}(\phi_n,\xi|\psi)+\tilde{O}(\phi_n,-\xi|\psi))$ and $O^{\rm w}_{\mathcal{I}}(\phi_n|\psi)=\frac{\hbar}{2\xi}(\tilde{O}(\phi_n,\xi|\psi)-\tilde{O}(\phi_n,-\xi|\psi))$.  

Note that, for a given fixed value of $\phi_n$, $\tilde{O}(\phi_n,\xi|\psi)$ in general fluctuates randomly due to the fluctuations of $\xi$. Averaging $\tilde{O}(\phi_n,\xi|\psi)$ over $\xi$, we obtain, by virtue of Eq. (\ref{Planck constant}), the real part of the weak value, i.e. 
\begin{eqnarray}
\int{\rm d}\xi\tilde{O}(\phi_n,\xi|\psi)\chi(\xi)=O^{\rm w}_{\mathcal{R}}(\phi_n|\psi).
\label{real part of weak value as the conditional average of c-valued physical quantity}
\end{eqnarray}
Hence, $O^{\rm w}_{\mathcal{R}}(\phi_n|\psi)$ can be seen as the conditional average of $\tilde{O}$ given $\phi_n$. Moreover, it is also clear from Eq. (\ref{real and imaginary parts of weak value vs compatibility and incompatibility}) that the fluctuations of $\tilde{O}(\phi_n,\xi|\psi)$ about its conditional average $O^{\rm w}_{\mathcal{R}}(\phi_n|\psi)$, i.e., the fluctuations of the error term $\frac{\xi}{\hbar}O^{\rm w}_{\mathcal{I}}(\phi_n|\psi)$ in Eq. (\ref{fundamental epistemic decomposition}), becomes stronger as the incompatibility between $\hat{O}$ and $\hat{\Pi}_{\phi_n}$ over $\ket{\psi}$ is larger. An epistemic interpretation of the two decomposing terms on the right-hand side of Eq. (\ref{fundamental epistemic decomposition}) within an operational scheme of optimal estimation will be suggested in Sec. \ref{Optimal estimation under epistemic restriction}. 

For an arbitrary pre-selected state $\ket{\psi}$ and a reference coordinate basis $\{\ket{\phi_n}\}$, we further assume that the coordinate value $\phi_n$ is distributed according to the Born's rule. Namely, the probability that $\phi_n$ is sampled is given by ${\rm Pr}(\phi_n|\psi)=\big|\braket{\phi_n|\psi}\big|^2$. Moreover, we define the ensemble average of a c-valued function $f(\phi_n,\xi|\psi)$ over the joint probability distribution ${\rm Pr}(\phi_n,\xi|\psi)=|\braket{\phi_n|\psi}|^2\chi(\xi)$ of $(\phi_n,\xi)$ as in the conventional probability theory, i.e., $\braket{f(\phi_n,\xi|\psi)}\doteq\sum_n\int{\rm d}\xi f(\phi_n,\xi|\psi)\chi(\xi)\big|\braket{\phi_n|\psi}\big|^2$. It is then straightforward to show that the average of $\tilde{O}(\phi_n,\xi|\psi)$ defined in Eq. (\ref{fundamental epistemic decomposition}) over the distribution of $(\phi_n,\xi)$ is given by the real part of the quantum expectation value of $\hat{O}$ over $\ket{\psi}$, i.e., 
\begin{eqnarray}
&&\braket{\tilde{O}(\phi_n,\xi|\psi)}\nonumber\\
&=&\sum_n\int{\rm d}\xi\big(O^{\rm w}_{\mathcal{R}}(\phi_n|\psi)+\frac{\xi}{\hbar}O^{\rm w}_{\mathcal{I}}(\phi_n|\psi)\big)\chi(\xi)|\braket{\phi_n|\psi}|^2\nonumber\\
&=&\sum_nO^{\rm w}_{\mathcal{R}}(\phi_n|\psi)|\braket{\phi_n|\psi}|^2\doteq\braket{O^{\rm w}_{\mathcal{R}}(\phi_n|\psi)}\nonumber\\
&=&{\rm Re}\big\{\braket{\psi|\hat{O}|\psi}\big\}=\frac{1}{2}\braket{\psi|(\hat{O}+\hat{O}^{\dagger})|\psi},
\label{real part of quantum expectation value as mean value of real-global-stochastic variable}
\end{eqnarray}
where we have used Eqs. (\ref{real and imaginary parts of weak value vs compatibility and incompatibility}) and (\ref{Planck constant}), and the completeness of the basis $\{\ket{\phi_n}\}$. We also have, using again Eqs. (\ref{real and imaginary parts of weak value vs compatibility and incompatibility}), (\ref{fundamental epistemic decomposition}) and (\ref{Planck constant}),  
\begin{eqnarray}
&&\Big\langle\frac{\xi}{\hbar}\tilde{O}(\phi_n,\xi|\psi)\Big\rangle\nonumber\\
&=&\sum_n\int{\rm d}\xi\Big(\frac{\xi}{\hbar}O^{\rm w}_{\mathcal{R}}(\phi_n|\psi)+\frac{\xi^2}{\hbar^2}O^{\rm w}_{\mathcal{I}}(\phi_n|\psi)\Big)\chi(\xi)|\braket{\phi_n|\psi}|^2\nonumber\\
&=&{\rm Im}\big\{\braket{\psi|\hat{O}|\psi}\big\}=\frac{1}{2i}\braket{\psi|(\hat{O}-\hat{O}^{\dagger})|\psi}.
\label{imaginary part of quantum expectation value as mean value of real-global-stochastic variable}
\end{eqnarray}

From Eqs. (\ref{real part of quantum expectation value as mean value of real-global-stochastic variable}) and (\ref{imaginary part of quantum expectation value as mean value of real-global-stochastic variable}), the expectation value of any quantum operator $\hat{O}$ over an arbitrary pre-selected state $\ket{\psi}$ can thus be decomposed as, 
\begin{eqnarray}
\braket{\tilde{O}(\phi_n,\xi|\psi)}+i\Big\langle\frac{\xi}{\hbar}\tilde{O}(\phi_n,\xi|\psi)\Big\rangle=\braket{\psi|\hat{O}|\psi}. 
\label{quantum expectation value as mean value of real-global-stochastic variable: general operators}
\end{eqnarray}
Hence, it can be seen as the ensemble average of a complex c-valued variable defined as: $\tilde{O}(\phi_n,\xi|\psi)+i\frac{\xi}{\hbar}\tilde{O}(\phi_n,\xi|\psi)$. Observe that the two terms on the left-hand side of Eq. (\ref{quantum expectation value as mean value of real-global-stochastic variable: general operators}) depends on the choice of basis $\{\ket{\phi_n}\}$, while right-hand side does not. Moreover, if the operator $\hat{O}$ is Hermitian, the second term on the left-hand side of Eq. (\ref{quantum expectation value as mean value of real-global-stochastic variable: general operators}) vanishes so that we have  
\begin{eqnarray}
\braket{\tilde{O}(\phi_n,\xi|\psi)}=\braket{\psi|\hat{O}|\psi}.
\label{quantum expectation value of Hermitian operator as the average value of c-valued physical quantity}
\end{eqnarray}
As a comparison, the quantum expectation value of $\hat{O}$ over $\ket{\psi}$ can be expressed directly in terms of the average of weak value as $\braket{\psi|\hat{O}|\psi}=\sum_n O^{\rm w}(\phi_n|\psi)|\braket{\phi_n|\psi}|^2$. Note however that $O^{\rm w}(\phi_n|\psi)$ is in general complex-valued even when $\hat{O}$ is Hermitian. In contrast to this, in Eq. (\ref{quantum expectation value of Hermitian operator as the average value of c-valued physical quantity}), we have expressed the quantum expectation value of a Hermitian operator in terms of the statistical average of a real-valued variable at the cost of introducing an auxiliary variable $\xi$. 

Several important properties of the above class of c-valued physical quantities are instructive for later reference:

{\it Correspondence with the strong value}. Assume that the reference coordinate basis is given by the complete set of eigenvectors $\{\ket{o_n}\}$ of a Hermitian quantum observable $\hat{O}$ with the set of real eigenvalues $\{o_n\}$. In this case, $[\hat{\Pi}_{o_n},\hat{O}]=0$, so that from Eq. (\ref{real and imaginary parts of weak value vs compatibility and incompatibility}) we have $O^{\rm w}_{\mathcal{I}}(o_n|\psi)=0$. Accordingly, the $\xi$-dependent error term on the right-hand side of Eq. (\ref{fundamental epistemic decomposition}) vanishes. Moreover, the real part of the weak value is given by $O^{\rm w}_{\mathcal{R}}(o_n|\psi)=o_n$, i.e., the eigenvalue of $\hat{O}$ associated with the eigenvector $\ket{o_n}$. One thus has
\begin{eqnarray}
\tilde{O}(o_n,\xi|\psi)=O^{\rm w}_{\mathcal{R}}(o_n|\psi)=o_n. 
\label{deterministic c-valued for coordinate given by its complete set of eigenvector}
\end{eqnarray}
Hence, regardless the value of $\xi$, the c-valued physical quantity is equal to the real eigenvalue $o_n$ of $\hat{O}$, the standard strong value obtained in strong measurement. This is also the case when $\ket{\psi}$ is given by one of the eigenstates of $\hat{O}$ for arbitrary choices of basis. In this sense, $\tilde{O}(\phi_n,\xi|\psi)$ may be regarded as a generalization of the strong value. Recall that strong value associated with $\hat{O}$ is definite only when the pre-selected state $\ket{\psi}$ is given by one of the eigenstates of $\hat{O}$; otherwise it is indefinite when $\ket{\psi}$ is a superposition of different eigenstates of $\hat{O}$. By contrast, the c-valued physical quantity $\tilde{O}(\phi_n,\xi|\psi)$ is always definite for all $\ket{\psi}$. Hence, by introducing $\xi$, we have somehow transformed the quantum coherence-induced indefiniteness in the strong value, into a fluctuation of a definite c-valued physical quantity. Note further that even when $\hat{O}$ is nonHermitian, $\tilde{O}(\phi_n,\xi|\psi)$ defined in Eq. (\ref{fundamental epistemic decomposition}) is always real. However, in this case, even when $\ket{\phi_n}$ or $\ket{\psi}$ is the eigenstate of $\hat{O}$, since the associated eigenvalue may be complex, $O^{\rm w}_{\mathcal{I}}(\phi_n|\psi)$ is in general not vanishing so that $\tilde{O}(\phi_n,\xi|\psi)$ may still fluctuate randomly due to the fluctuation of $\xi$. Hence, if we wish $\tilde{O}(\phi_n,\xi|\psi)$ defined in Eq. (\ref{fundamental epistemic decomposition}) to be independent of $\xi$ when $\ket{\phi_n}$ or $\ket{\psi}$ is the eigenstate of $\hat{O}$, then $\hat{O}$ must be Hermitian.  

{\it Complementarity between the error terms.} Consider two noncommuting quantum observables, $\hat{A}$ and $\hat{B}$, $[\hat{A},\hat{B}]\neq 0$, so that there is no common complete set of vectors in which $\hat{A}$ and $\hat{B}$ are jointly diagonalized. Now, take the complete set of eigenvectors $\{\ket{a_n}\}$ of $\hat{A}$ with the associated set of eigenvalues $\{a_n\}$ as the reference coordinate basis. As shown above, we thus have $\tilde{A}(a_n,\xi|\psi)=a_n$, i.e., it is independent of the global random variable $\xi$. However, observe that, in this case, in general we have $[\hat{\Pi}_{a_n},\hat{B}]\neq 0$ so that $B^{\rm w}_{\mathcal{I}}\neq 0$. Accordingly, $\tilde{B}(a_n,\xi|\psi)$ must in general fluctuate due to the fluctuation of $\xi$ via the $\xi$-dependent error term. Moreover, this fluctuation is stronger as the incompatibility between $\hat{B}$ and $\hat{\Pi}_{a_n}$ over $\ket{\psi}$ is larger. Hence, for an arbitrary pair of noncommuting quantum observables, if one chooses a reference coordinate basis so that the error term of the c-valued physical quantity associated with one of the observables is vanishing, the error term of the c-valued physical quantity associated with the other observable is in general not vanishing, and vice versa. This complementarity between the error terms of the c-valued physical quantities associated with two noncommuting observables thus captures the complementarity between the randomness of the quantum strong values.     

{\it Violation of product rule}. Like the weak value, the c-valued physical quantities defined in Eq. (\ref{fundamental epistemic decomposition}) conserve the additive structure of the quantum operators. Namely, given $\hat{C}=\hat{A}+\hat{B}$, we have $\tilde{C}(\phi_n,\xi|\psi)=\tilde{A}(\phi_n,\xi|\psi)+\tilde{B}(\phi_n,\xi|\psi)$. However, for $\hat{C}=\hat{A}\hat{B}$, even when $\hat{A}$ and $\hat{B}$ are commuting, one in general has 
\begin{eqnarray}
\tilde{C}(\phi_n,\xi|\psi)\neq\tilde{A}(\phi_n,\xi|\psi)\tilde{B}(\phi_n,\xi|\psi). 
\label{contextual c-valued physical quantities: does not conserve functional relationship of operators}
\end{eqnarray}
Hence, it in general does not conserve the functional relationship of the original operators. The equality between the left and the right-hand sides of Eq. (\ref{contextual c-valued physical quantities: does not conserve functional relationship of operators}) applies when the pre-selected state $\ket{\psi}$ or the post-selected state $\ket{\phi_n}$ is the joint eigenstate of the two Hermitian operators $\hat{A}$ and $\hat{B}$. As an illustrative example, consider the quantum Hamiltonian of a free particle with mass $m$ in one spatial dimension: $\hat{H}=\hat{p}^2/2m$. Taking the complete set of the eigenvectors $\{\ket{q}\}$ of the position operator $\hat{q}$ as the reference coordinate basis, one has 
\begin{eqnarray}
\tilde{H}(q,\xi|\psi)&=&\frac{(\partial_qS)^2}{2m}-\frac{\hbar^2}{2m}\frac{\partial_q^2\sqrt{\rho}}{\sqrt{\rho}}-\frac{\xi}{2\rho}\partial_q\Big(\rho\frac{\partial_qS}{m}\Big)\nonumber\\
&\neq&\frac{(\partial_qS-\frac{\xi}{2}\frac{\partial_q\rho}{\rho})^2}{2m}=\frac{\tilde{p}(q,\xi|\psi)^2}{2m},  
\label{contextual c-valued physical quantities: energy - momentum relationship}
\end{eqnarray}
where we have expressed the wave function in the polar form, i.e., $\psi(q)\doteq\braket{q|\psi}=\sqrt{\rho(q)}e^{iS(q)/\hbar}$, so that $S(q)=\hbar{\rm Arg}\{\psi(q)\}$, and $\rho(q)=|\psi(q)|^2$. In particular, while the right-hand side of Eq. (\ref{contextual c-valued physical quantities: energy - momentum relationship}) is always non-negative, the left-hand side may take negative value. As an example of such a negative c-valued kinetic energy, consider the case when the pre-selected state is given by a Gaussian wave function $\psi(q)\sim e^{-q^2/4\sigma_q^2}$. Then, one has $\tilde{H}(q,\xi|\psi)=-\frac{\hbar^2}{2m}(-\frac{1}{2\sigma_q^2}+\frac{q^2}{4\sigma_q^4})$, so that $\tilde{H}(q,\xi|\psi)\le 0$ for $q^2\ge 2\sigma_q^2$. Interestingly, as will be shown in Sec. \ref{c-valued physical quantities as deterministic hidden variable models for quantum expectation values of products operators}, the ensemble average of both sides in Eq. (\ref{contextual c-valued physical quantities: does not conserve functional relationship of operators}) are always equal. For weak values, the above feature for the product of operators is the key to many quantum paradoxes involving weak measurement with post-selection \cite{Ahoronov-Rohrlich book}.  

\subsection{C-valued physical quantities as a real-deterministic contextual hidden variable model for a class of quantum expectation values \label{c-valued physical quantities as deterministic hidden variable models for quantum expectation values of products operators}}

We first obtain the following mathematical result for the average of the product of two c-valued physical quantities. Consider an arbitrary pair of operators $\hat{A}$ and $\hat{B}$. For any pre-selected state $\ket{\psi}$ and a reference coordinate basis $\{\ket{\phi_n}\}$, the ensemble average of the product between $\tilde{A}(\phi_n,\xi|\psi)$ associated with $\hat{A}$, and $\tilde{B}(\phi_n,\xi|\psi)$ associated with $\hat{B}$, is equal to the quantum expectation value of the symmetrized product of the two operators. Namely, using Eqs. (\ref{fundamental epistemic decomposition}) and (\ref{Planck constant}), we straightforwardly obtain 
\begin{eqnarray}
&&\braket{\tilde{A}(\phi_n,\xi|\psi)\tilde{B}(\phi_n,\xi|\psi)}\nonumber\\
&=&\sum_n\big(A^{\rm w}_{\mathcal{R}}(\phi_n|\psi)B^{\rm w}_{\mathcal{R}}(\phi_n|\psi)+\frac{\overline{\xi^2}}{\hbar^2}A^{\rm w}_{\mathcal{I}}(\phi_n|\psi)B^{\rm w}_{\mathcal{I}}(\phi_n|\psi)\big)\nonumber\\
&&\times\big|\braket{\phi_n|\psi}\big|^2\nonumber\\
&=&\sum_n{\rm Re}\Big\{\frac{\braket{\phi_n|\hat{A}|\psi}}{\braket{\phi_n|\psi}}\frac{\braket{\phi_n|\hat{B}|\psi}^*}{\braket{\phi_n|\psi}^*}\Big\}
\big|\braket{\phi_n|\psi}\big|^2\nonumber\\
&=&\braket{\psi|(\hat{A}^{\dagger}\hat{B}+\hat{B}^{\dagger}\hat{A})/2|\psi}.
\label{real-global-stochastic representation of quantum anticommutator: general operators}
\end{eqnarray}
Here, to get the second equality, we have used Eq. (\ref{Planck constant}) and a simple mathematical identity ${\rm Re}\{ab^*\}={\rm Re}\{a\}{\rm Re}\{b\}+{\rm Im}\{a\}{\rm Im}\{b\}$ for two arbitrary complex numbers $a$ and $b$. One can check that when $\hat{B}=\hat{1}$, we regain Eq. (\ref{real part of quantum expectation value as mean value of real-global-stochastic variable}). Note that while the decomposition in the second line of Eq. (\ref{real-global-stochastic representation of quantum anticommutator: general operators}) depends on the choice of the coordinate basis $\{\ket{\phi_n}\}$, the last line (i.e., the right-hand side) is basis-independent.  

As a corollary of Eq. (\ref{real-global-stochastic representation of quantum anticommutator: general operators}), when both $\hat{A}$ and $\hat{B}$ are Hermitian, we get 
\begin{eqnarray}
\braket{\tilde{A}(\phi_n,\xi|\psi)\tilde{B}(\phi_n,\xi|\psi)}=\braket{\psi|(\hat{A}\hat{B}+\hat{B}\hat{A})/2|\psi}.
\label{real-global-stochastic representation of quantum anticommutator}
\end{eqnarray}
Moreover, if the two observables are commuting, which is e.g. the case when the two quantum observables operate locally on two spatially separated systems, one has 
\begin{eqnarray}
\braket{\tilde{A}(\phi_n,\xi|\psi)\tilde{B}(\phi_n,\xi|\psi)}=\braket{\psi|\hat{A}\hat{B}|\psi}. 
\label{real-global-stochastic representation of quantum correlation: commuting observables}
\end{eqnarray}
Next, taking $\hat{B}=\hat{A}$ in Eq. (\ref{real-global-stochastic representation of quantum anticommutator: general operators}), one obtains
\begin{eqnarray}
\braket{\tilde{A}(\phi_n,\xi|\psi)^2}=\braket{\psi|\hat{A}^{\dagger}\hat{A}|\psi}.
\label{real-global-stochastic representation of quantum variance: general operators}
\end{eqnarray}
As an example of the above result, taking the ensemble average of both sides of Eq. (\ref{contextual c-valued physical quantities: energy - momentum relationship}), one has 
\begin{eqnarray}
&&\braket{\tilde{H}(q,\xi|\psi)}=\braket{\psi|\hat{H}|\psi}\nonumber\\
&=&\braket{\psi|\hat{p}^2/2m|\psi}=\braket{\tilde{p}(q,\xi|\psi)^2/2m}, 
\label{contextual kinetic energy via weak value}
\end{eqnarray}
where the first and the last equalities are due to respectively Eqs. (\ref{quantum expectation value of Hermitian operator as the average value of c-valued physical quantity}) and (\ref{real-global-stochastic representation of quantum variance: general operators}). Note that the equality between the first and the last ensemble averages in Eq. (\ref{contextual kinetic energy via weak value}) can be checked directly by noting Eq. (\ref{Planck constant}) and using the identity $\frac{1}{4}\big(\frac{\partial_q\rho}{\rho}\big)^2=-\frac{\partial_q^2\sqrt{\rho}}{\sqrt{\rho}}+\frac{1}{2}\frac{\partial_q^2\rho}{\rho}$, where $\rho(q)=|\braket{q|\psi}|^2$, and assuming that $\rho(q)$ is vanishing at the boundary.   

Observe that the cross terms, i.e., $A^{\rm w}_{\mathcal{R}(\mathcal{I})}(\phi_n|\psi)B^{\rm w}_{\mathcal{I}(\mathcal{R})}(\phi_n|\psi)$, did not appear in the second line of Eq. (\ref{real-global-stochastic representation of quantum anticommutator: general operators}) because of the vanishing average of $\xi$ assumed in Eq. (\ref{Planck constant}). Moreover, we emphasize that the globalness (i.e., the nonseparability) of $\xi$ is indeed in general indispensable when the operators $\hat{A}$ and $\hat{B}$ act locally on two spatially separated systems. Otherwise, if $\xi$ were separable into two independent random variables each belonging to the two systems, i.e., $\xi=(\xi_A,\xi_B)$ so that 
\begin{eqnarray}
\chi(\xi)=\chi_A(\xi_A)\chi_B(\xi_B),~~{\rm with}~~\overline{\xi}_A=0=\overline{\xi}_B, 
\label{non-global random variable}
\end{eqnarray}
then, we have $\overline{\xi_A\xi_B}=\overline{\xi_A}\times\overline{\xi_B}=0$ so that the second term in the second line of Eq. (\ref{real-global-stochastic representation of quantum anticommutator: general operators}) vanishes. Further elaboration of this fact and its deep relation with quantum entanglement will be discussed somewhere else. 

As a direct implication of Eq. (\ref{real-global-stochastic representation of quantum anticommutator}), the quantum covariance of two Hermitian observables $\hat{A}$ and $\hat{B}$ over a quantum state $\ket{\psi}$ can be expressed as the classical covariance of the associated c-valued physical quantities $\tilde{A}(\phi_n,\xi|\psi)$ and $\tilde{B}(\phi_n,\xi|\psi)$, i.e., 
\begin{eqnarray}
C_{\tilde{A}\tilde{B}}[\psi]&\doteq&\braket{\tilde{A}(\phi_n,\xi|\psi)\tilde{B}(\phi_n,\xi|\psi)}\nonumber\\
&&-\braket{\tilde{A}(\phi_n,\xi|\psi)}\braket{\tilde{B}(\phi_n,\xi|\psi)}\nonumber\\
&=&\frac{1}{2}\braket{\psi|\big(\hat{A}\hat{B}+\hat{B}\hat{A}\big)|\psi}-\braket{\psi|\hat{A}|\psi}\braket{\psi|\hat{B}|\psi}\nonumber\\
&=&C_{\hat{A}\hat{B}}[\psi]. 
\label{ER classical covariance is equal to quantum covariance}
\end{eqnarray}
Taking $\hat{A}=\hat{B}=\hat{O}$ in Eq. (\ref{ER classical covariance is equal to quantum covariance}), the quantum variance of an observable $\hat{O}$ over a quantum state $\ket{\psi}$ can thus be expressed as the classical variance of the associated c-valued physical quantity $\tilde{O}(n,\xi|\psi)$, i.e., 
\begin{eqnarray}
\sigma_{\tilde{O}}^2[\psi]&\doteq&\braket{\tilde{O}(\phi_n,\xi|\psi)^2}-\braket{\tilde{O}(\phi_n,\xi|\psi)}^2\nonumber\\
&=&\braket{\psi|\hat{O}^2|\psi}-\braket{\psi|\hat{O}|\psi}^2\nonumber\\
&=&\sigma_{\hat{O}}^2[\psi].
\label{ER classical variance is equal to quantum variance}  
\end{eqnarray}

Moreover, using again Eq. (\ref{real-global-stochastic representation of quantum anticommutator}), the statistical deviation (it is mistakenly written as statistical distance in the version published in Phys. Rev. A 103, 022215 (2021) which has a completely different meaning) between two quantum observables $\hat{A}$ and $\hat{B}$ over a quantum state $\ket{\psi}$ can be expressed as the $l_2$-norm of the associated c-valued physical quantities, i.e.:
\begin{eqnarray}
&&\big\langle\big(\tilde{A}(\phi_n,\xi|\psi)-\tilde{B}(\phi_n,\xi|\psi)\big)^2\big\rangle\nonumber\\
&=&\langle(\tilde{A}(\phi_n,\xi|\psi)^2-2\tilde{A}(\phi_n,\xi|\psi)\tilde{B}(\phi_n,\xi|\psi)+\tilde{B}(\phi_n,\xi|\psi)^2)\rangle\nonumber\\
&=&\braket{\psi|\big(\hat{A}^2-2\frac{(\hat{A}\hat{B}+\hat{B}\hat{A})}{2}+\hat{B}^2\big)|\psi}\nonumber\\
&=&\braket{\psi|(\hat{A}-\hat{B})^2|\psi}. 
\label{l_2 norm between two c-valued physical quantities is equal to the statistical deviation between operators}
\end{eqnarray}
The statistical deviation between two observables plays a central role in the derivations of the error-disturbance and joint-measurement uncertainty relations \cite{Weston error-disturbance UR,Ozawa error-disturbance UR,Branciard error-disturbance UR,Hall exact UR,Hall prior information}. We shall show in Sec. \ref{Optimal estimation under epistemic restriction} that Eq. (\ref{l_2 norm between two c-valued physical quantities is equal to the statistical deviation between operators}) provides a mathematical link between our formalism and that of Johansen \cite{Johansen weak value best estimation} and Hall \cite{Hall exact UR,Hall prior information} with regard to the interpretation of the decomposition of the c-valued physical quantities defined in Eq. (\ref{fundamental epistemic decomposition}) as describing an optimal scheme of estimation. Note further that, since $\hat{A}$ and $\hat{B}$ are in general noncommuting so that they cannot be jointly measured, the right-hand side of Eq. (\ref{l_2 norm between two c-valued physical quantities is equal to the statistical deviation between operators}) does not in general have a clear operational meaning. In this sense, the left-hand side can be seen as an operational definition of the right-hand side in terms of c-valued physical quantities which can be obtained by using weak measurement with post-selection. 

From Eqs. (\ref{real-global-stochastic representation of quantum anticommutator: general operators}) and (\ref{deterministic c-valued for coordinate given by its complete set of eigenvector}), we also obtain the following result. Consider two arbitrary operators $\hat{A}$, $\hat{B}$, and a function $f(\hat{C})$ of a Hermitian operator $\hat{C}$. Taking the complete set of eigenstates $\{\ket{c_n}\}$ of $\hat{C}$ as the reference coordinate basis to define the c-valued physical quantities, we obtain 
\begin{eqnarray}
&&\braket{\psi|\big(\frac{1}{2}(\hat{A}^{\dagger}\hat{B}+\hat{B}^{\dagger}\hat{A})+f(\hat{C})\big)|\psi}\nonumber\\
&=&\braket{\tilde{A}(c_n,\xi|\psi)\tilde{B}(c_n,\xi|\psi)+f(\tilde{C}(c_n,\xi|\psi))},\nonumber\\
&=&\braket{\tilde{A}(c_n,\xi|\psi)\tilde{B}(c_n,\xi|\psi)+f(c_n)},
\label{generalized equivalence theorem}
\end{eqnarray}
where in the last line we have used $\tilde{C}(c_n,\xi|\psi)=c_n$. As a specific example, for $\hat{A}=\hat{B}=\hat{p}$ and $C=\hat{q}$, we have $\braket{\psi|\big(\frac{1}{2}\hat{p}^2+f(\hat{q})\big)|\psi}=\braket{\frac{1}{2}\tilde{p}(q,\xi|\psi)^2+f(q)}$. Finally, assume further that $\xi$ satisfies $\overline{\xi^3}=0$. Then we obtain, for an arbitrary pair of operators $\hat{A}$ and $\hat{B}$, 
\begin{eqnarray}
\braket{\psi|\hat{A}^{\dagger}\hat{B}|\psi}&=&\braket{\tilde{A}(\phi_n,\xi|\psi)\tilde{B}(\phi_n,\xi|\psi)}\nonumber\\
&+&i\Big\langle\frac{\xi}{\hbar}\tilde{A}(\phi_n,-\xi|\psi)\tilde{B}(\phi_n,\xi|\psi)\Big\rangle.
\label{real-global-stochastic representation of quantum correlation}
\end{eqnarray}
See Appendix \ref{C-valued representation of quantum expectation value of product of two operators} for a proof. 

All the above results show that the class of c-valued physical quantities defined in Eq. (\ref{fundamental epistemic decomposition}) provides a real-deterministic hidden variable model for quantum expectation value of a certain class of operators. Note that Hosoya {\it et al}. in \cite{Hosoya-Shikano counterfactual value} and Hall {\it et al}. in \cite{Hall weak value to quantum uncertainty} also obtained a representation of the product of two arbitrary operators in terms of the average of weak values. Indeed inserting Eq. (\ref{fundamental epistemic decomposition}) into Eq. (\ref{real-global-stochastic representation of quantum correlation}) we regain their result: 
\begin{eqnarray}
\braket{\psi|\hat{A}^{\dagger}\hat{B}|\psi}=\sum_n A^{\rm w}(\phi_n|\psi)^*B^{\rm w}(\phi_n|\psi)\big|\braket{\phi_n|\psi}\big|^2. 
\end{eqnarray}
However, in their representation, the expectation value of the product of two operators are reconstructed directly from the average of the product of the complex weak values even when the operators $\hat{A}$ and $\hat{B}$ are Hermitian, i.e., the hidden variables are given by complex-valued variables. By contrast, within our model, as can be seen in Eqs. (\ref{real-global-stochastic representation of quantum anticommutator: general operators}) or (\ref{real-global-stochastic representation of quantum correlation}), they are reconstructed from the c-valued physical quantities which are always real. This is made possible via the introduction of a global random variable $\xi$. The real-valued physical quantities arguably give a more direct and transparent correspondence with the standard strong value of a Hermitian observable given by the real eigenvalue. As will be discussed at the end of Sec. \ref{Optimal estimation under epistemic restriction}, it also offers a smooth classical-correspondence.  Moreover, by introducing $\xi$ in the definition of the c-valued physical quantities of Eq. (\ref{fundamental epistemic decomposition}), we have been able to isolate the commutator and anticommutator between the operator and the coordinate basis, and transform them into two different kinds of real-valued fluctuations, which is otherwise concealed in the complex weak value. We show in Sec. \ref{Quantum uncertainty relation from real deterministic c-valued physical quantities} that this decomposition gives a fresh insight into the physical origin of the Kennard-Robertson-Schr\"odinger uncertainty relation. 

It is important to note that the reconstruction of the quantum expectation value of certain class of operators in terms of the ensemble average of real-deterministic hidden variables given by the c-valued physical quantities defined in Eq. (\ref{fundamental epistemic decomposition}) is contextual in the following sense. First, the definition of c-valued physical quantities depends on the choice of the reference coordinate basis $\{\ket{\phi_n}\}$. Since each reference basis has a distinct operational meaning in terms of post-selection in the weak value measurement, it thus provides a measurement context to the value assignment of the c-valued physical quantities. A similar post-selection measurement context has been discussed in detailed in Ref. \cite{Hofmann contextual value - weak value} for the reconstruction of quantum variance of an observable in terms of fluctuations of complex weak values. 

Next, consider again the reconstruction of the quantum expectation value of kinetic energy $\braket{\psi|\hat{p}^2/2m|\psi}$ via the above hidden variable model using c-valued physical quantities by taking the position eigenstates $\{\ket{q}\}$ as the reference coordinate basis. Equation (\ref{contextual kinetic energy via weak value}) shows that there are two ways to achieve this, i.e., by directly averaging the c-valued kinetic energy $\tilde{H}(q,\xi|\psi)$ or averaging $\tilde{p}(q,\xi|\psi)^2/2m$ computed from the c-valued momentum $\tilde{p}(q,\xi|\psi)$. Note however that while their averages are equal, as shown in Eq. (\ref{contextual c-valued physical quantities: energy - momentum relationship}), the two quantities are not, i.e., $\tilde{H}(q,\xi|\psi)\neq \tilde{p}(q,\xi|\psi)^2/2m$. Moreover, they are operationally obtained in two different schemes of weak measurement, i.e., weak kinetic energy measurement and weak momentum measurement, respectively. The above observation generalizes to an arbitrary quantum operator that can be written as a product of two operators, $\hat{C}=\hat{A}^{\dagger}\hat{B}$. Namely, the quantum expectation value of $\hat{C}$ over $\ket{\psi}$, can be reconstructed from the ensemble average of c-valued physical quantity $\tilde{C}(\phi_n,\xi|\psi)$ via Eq. (\ref{quantum expectation value as mean value of real-global-stochastic variable: general operators}), or from the ensemble average of the combination of c-valued physical quantities $\tilde{A}(\phi_n,\xi|\psi)$ and $\tilde{B}(\phi_n,\xi|\psi)$ via Eq. (\ref{real-global-stochastic representation of quantum correlation}). This observation shows that if the operator can be written as a product of two operators, the choice of weak measurement defines a particular measurement context, all leading to the same expectation value. 

Finally, let us consider a mixed state or density matrix $\hat{\varrho}$ which can be decomposed in terms of a set of pure states as $\hat{\varrho}=\sum_{\mu}{\rm Pr}(\psi_{\mu})\ket{\psi_{\mu}}\bra{\psi_{\mu}}$. It can be prepared e.g. by sampling the pure state $\ket{\psi_{\mu}}$ with a probability ${\rm Pr}(\psi_{\mu})$ and deleting the record about the index $\mu$. For a coordinate basis $\{\ket{\phi_n}\}$, the preparation thus samples $\tilde{O}(\phi_n,\xi|\psi_{\mu})$ with a joint probability ${\rm Pr}(\psi_{\mu},\phi_n)\doteq{\rm Pr}(\psi_{\mu})|\braket{\phi_n|\psi_{\mu}}|^2$. Noting this, the results presented so far for pure states can then be generalized to the above statistical mixture of c-valued physical quantities by using the Bayes rule in conventional probability theory. As an example, computing the correlation between the mixture of two c-valued physical quantities $\tilde{A}$ and $\tilde{B}$ we obtain:
\begin{eqnarray}
&&\sum_{\mu}{\rm Pr}(\psi_{\mu})\braket{\tilde{A}(\phi_n,\xi|\psi_{\mu})\tilde{B}(\phi_n,\xi|\psi_{\mu})}\nonumber\\
&=&\sum_{\mu}{\rm Pr}(\psi_{\mu})\braket{\psi_{\mu}|(\hat{A}\hat{B}^{\dagger}+\hat{B}^{\dagger}\hat{A})/2|\psi_{\mu}}\nonumber\\
&=&{\rm Tr}\{\hat{\varrho}(\hat{A}\hat{B}^{\dagger}+\hat{B}^{\dagger}\hat{A})/2\},
\end{eqnarray}
where we have used Eq. (\ref{real-global-stochastic representation of quantum anticommutator: general operators}).  

A couple of notes are in order. First, the extension to mixture of pure states above does not employ the usual definition of the weak value for the pre-selected mixed state $\hat{\varrho}$, which for an operator $\hat{O}$ and a post-selected state $\ket{\phi_n}$ is defined as  \cite{Hall prior information,Johansen weak value best estimation,Wiseman generalized weak value for mixed state}
\begin{eqnarray}
O^{\rm w}(\phi_n|\varrho)\doteq\frac{{\rm Tr}\{\hat{\Pi}_{\phi_n}\hat{O}\hat{\varrho}\}}{{\rm Tr}\{\hat{\Pi}_{\phi_n}\hat{\varrho}\}}. 
\label{weak value for general pre-selected mixed state}
\end{eqnarray}
Second, and importantly, recall that a mixed state can be decomposed as a mixture of pure states in various different ways. Noting this, the set of c-valued physical quantities associated with a mixed state clearly depends on the specific form of the decomposition of the mixed state in terms of the pure states. Namely, different schemes for the preparation of $\hat{\varrho}$ as a mixture of pure states $\hat{\varrho}\doteq\sum_{\mu}{\rm Pr}(\psi_{\mu})\ket{\psi_{\mu}}\bra{\psi_{\mu}}$, leads to different mixture of set of c-valued physical quantities $\{\tilde{O}(\phi_n,\xi|\psi_{\mu})\}$ with the conditional probabilities ${\rm Pr}(\psi_{\mu}|\phi_n)\doteq\frac{{\rm Pr}(\psi_\mu)|\braket{\phi_n|\psi_{\mu}}|^2}{{\rm Tr}\{\hat{\Pi}_{\phi_n}\hat{\varrho}\}}$. In this sense, the hidden variable model of quantum expectation value with a mixed state using the c-valued physical quantities is preparation contextual. A similar preparation context for the reconstruction of quantum variance via the fluctuation of complex weak value is discussed in Ref. \cite{Hofmann contextual value - weak value} wherein the different decomposition of the mixed states are chosen via remote steering over an entangled pure state.

It is worth further noting however that upon averaging the c-valued physical quantity $\tilde{O}(\phi_n,\xi|\psi_{\mu})$ defined in Eq. (\ref{fundamental epistemic decomposition}) over the conditional probability ${\rm Pr}(\psi_{\mu}|\phi_n)=\frac{{\rm Pr}(\psi_\mu)|\braket{\phi_n|\psi_{\mu}}|^2}{{\rm Tr}\{\hat{\Pi}_{\phi_n}\hat{\varrho}\}}$, we obtain the c-valued physical quantity associated with the mixed state that is independent of its pure states decomposition, i.e.,
\begin{eqnarray}
\tilde{O}(\phi_n,\xi|\varrho)&\doteq&\sum_{\mu}\tilde{O}(\phi_n,\xi|\psi_{\mu}){\rm Pr}(\psi_{\mu}|\phi_n)\nonumber\\
&=&\sum_{\mu}\Big(O^{\rm w}_{\mathcal{R}}(\phi_n|\psi_{\mu})+\frac{\xi}{\hbar}O^{\rm w}_{\mathcal{I}}(\phi_n|\psi_{\mu})\Big)\nonumber\\
&&\times\frac{{\rm Pr}(\psi_\mu)|\braket{\phi_n|\psi_{\mu}}|^2}{{\rm Tr}\{\hat{\Pi}_{\phi_n}\hat{\varrho}\}}\nonumber\\
&=&O^{\rm w}_{\mathcal{R}}(\phi_n|\varrho)+\frac{\xi}{\hbar}O^{\rm w}_{\mathcal{I}}(\phi_n|\varrho),
\end{eqnarray}
where $O^{\rm w}_{\mathcal{R}}(\phi_n|\varrho)$ and $O^{\rm w}_{\mathcal{I}}(\phi_n|\varrho)$ are respectively the real and imaginary parts of weak value of $O^{\rm w}(\phi_n|\varrho)$ with the pre-selected mixed state $\hat{\varrho}$ defined in Eq. (\ref{weak value for general pre-selected mixed state}). 

\section{Quantum uncertainty relation as classical uncertainty relation between real-deterministic c-valued physical quantities \label{Quantum uncertainty relation from real deterministic c-valued physical quantities}}

Within the representation based on the c-valued physical quantities, given a pre-selected state, an arbitrary reference coordinate basis and $\xi$, all observables, including those of non-commuting set, are assigned deterministic (i.e., definite) real values. At first, this somehow contradicts the spirit of the complementarity principle that prohibits simultaneous value assignment to noncommuting observables. Nevertheless, we have shown that the c-valued physical quantities recover certain class of quantum statistics, including the quantum covariance of an observable and the statistical deviation between two observables. In this section, we shall rederive the KRS uncertainty relation directly from the real-deterministic c-valued physical quantities by imposing on them, being classical variables, to classical uncertainty relation.     

First, from Eq. (\ref{fundamental epistemic decomposition}), one can decompose the (classical) variance of the c-valued physical quantity $\tilde{O}(\phi_n,\xi|\psi)$ associated with a Hermitian operator $\hat{O}$, the pre-selected state $\ket{\psi}$, and an arbitrary reference coordinate basis $\{\ket{\phi_n}\}$, as
\begin{eqnarray}
\sigma_{\tilde{O}}^2[\psi]&=&\big\langle\big(\tilde{O}(\phi_n,\xi|\psi)-\braket{\tilde{O}(\phi_n,\xi|\psi)}\big)^2\big\rangle\nonumber\\
&=&\Big\langle\Big(\big[O^{\rm w}_{\mathcal{R}}(\phi_n|\psi)-\braket{O^{\rm w}_{\mathcal{R}}(\phi_n|\psi)}\big]+\frac{\xi}{\hbar}O^{\rm w}_{\mathcal{I}}(\phi_n|\psi)\Big)^2\Big\rangle\nonumber\\
&=&\Delta_{\tilde{O}}^2[\psi]+\mathcal{E}_{\tilde{O}}^2[\psi].
\label{decomposition of quantum uncertainty into classical-like part and genuine quantum part: A}
\end{eqnarray}
Here, we have used Eq. (\ref{Planck constant}) to get the last equality, and $\Delta_{\tilde{O}}^2[\psi]$ and $\mathcal{E}_{\tilde{O}}^2[\psi]$ for an arbitrary c-valued physical quantity $\tilde{O}(\phi_n,\xi|\psi)$ are defined as 
\begin{eqnarray}
\Delta_{\tilde{O}}^2[\psi]&\doteq&\big\langle\big(O^{\rm w}_{\mathcal{R}}(\phi_n|\psi)-\braket{O^{\rm w}_{\mathcal{R}}(\phi_n|\psi)}\big)^2\big\rangle,\nonumber\\
\mathcal{E}_{\tilde{O}}^2[\psi]&\doteq&\Big\langle\Big(\frac{\xi}{\hbar}O^{\rm w}_{\mathcal{I}}(\phi_n|\psi)-\Big\langle\frac{\xi}{\hbar}O^{\rm w}_{\mathcal{I}}(\phi_n|\psi)\Big\rangle\Big)^2\Big\rangle\nonumber\\
&=&\big\langle\big(O^{\rm w}_{\mathcal{I}}(\phi_n|\psi)\big)^2\big\rangle,  
\label{precision and accuracy}
\end{eqnarray}
where in the third line we have again used Eq. (\ref{Planck constant}). Hence, $\Delta_{\tilde{O}}^2[\psi]$ and $\mathcal{E}_{\tilde{O}}^2[\psi]$ are respectively the variance of the $\xi$-independent and $\xi$-dependent terms of the c-valued physical quantity $\tilde{O}(\phi_n,\xi|\psi)$, which are given by the variance of the real and imaginary parts of the weak value $O^{\rm w}(\phi_n|\psi)$.     

Upon inserting Eq. (\ref{decomposition of quantum uncertainty into classical-like part and genuine quantum part: A}) into Eq. (\ref{ER classical variance is equal to quantum variance}), we thus obtain the following general decomposition of quantum variance of a Hermitian operator $\hat{O}$ over a quantum state $\ket{\psi}$ into the variance of the $\xi$-independent term and the $\xi$-dependent error term of the associated c-valued physical quantity $\tilde{O}(\phi_n,\xi|\psi)$:
\begin{eqnarray}
\sigma_{\hat{O}}^2[\psi]=\sigma_{\tilde{O}}^2[\psi]=\Delta_{\tilde{O}}^2[\psi]+\mathcal{E}_{\tilde{O}}^2[\psi]. 
\label{decomposition of the quantum variance into the variance of the c-valued quantities}
\end{eqnarray}
Notice that the left-hand side of Eq. (\ref{decomposition of the quantum variance into the variance of the c-valued quantities}) is basis-independent while the decomposition on the right-hand side is basis-dependent. Hence, for a given quantum variance $\sigma_{\hat{O}}^2[\psi]$, by varying the choice of the basis $\{\ket{\phi_n}\}$ (i.e., the post-selection measurement context in the associated weak value measurement), we obtain a trade-off between the value of $\Delta_{\tilde{O}}^2[\psi]$ and $\mathcal{E}_{\tilde{O}}^2[\psi]$ on the right-hand side of Eq. (\ref{decomposition of the quantum variance into the variance of the c-valued quantities}), namely, between the strength of the fluctuations of the $\xi$-independent and the $\xi$-dependent error terms in the associated c-valued physical quantity $\tilde{O}(\phi_n,\xi|\psi)$. In particular, one can choose the complete set of eigenvectors $\{\ket{o_n}\}$ of $\hat{O}$ as the coordinate basis, so that the error term in $\tilde{O}(o_n,\xi|\psi)$ vanishes $\mathcal{E}_{\tilde{O}}^2[\psi]=0$ and we have $\sigma_{\hat{O}}^2[\psi]=\Delta_{\tilde{O}}^2[\psi]$. The decomposition of Eq. (\ref{decomposition of the quantum variance into the variance of the c-valued quantities}) can of course be proven directly by using the algebraic structure of the linear operators in Hilbert space \cite{Johansen weak value best estimation,Hall exact UR,Hall prior information}. Here, we have derived it by first mapping the triple $(\hat{O},\ket{\psi},\{\ket{\phi_n}\})$ into the c-valued physical quantity $\tilde{O}(\phi_n,\xi|\psi)$, and thus offering the interpretation of the quantum variance in terms of the fluctuation of a real-valued deterministic variable. This physical intuition is valuable in our derivation of and discussion about the KRS uncertainty relation below. 

Now, consider two arbitrary quantum observables $\hat{A}$ and $\hat{B}$. Let $\{\ket{b_n}\}$ be the complete set of eigenstates of $\hat{B}$ with the eigenvalues $\{b_n\}$, assumed for simplicity to be nondegenerate; hence, we have a spectral decomposition $\hat{B}=\sum_n b_n\ket{b_n}\bra{b_n}$. Let us take $\{\ket{b_n}\}$ as the reference coordinate basis providing the context to define the c-valued physical quantities. We thereby obtain, from Eq. (\ref{fundamental epistemic decomposition}),  
\begin{eqnarray}
\tilde{A}(b_n,\xi|\psi)&=&A^{\rm w}_{\mathcal{R}}(b_n|\psi)+\frac{\xi}{\hbar}A^{\rm w}_{\mathcal{I}}(b_n|\psi),~~{\rm and}\nonumber\\
\tilde{B}(b_n,\xi|\psi)&=&B^{\rm w}_{\mathcal{R}}(b_n|\psi)=b_n.  
\label{A and B from the eye of B}
\end{eqnarray}
Hence, using the reference basis $\{\ket{b_n}\}$, noting Eqs. (\ref{decomposition of quantum uncertainty into classical-like part and genuine quantum part: A}), (\ref{precision and accuracy}) and (\ref{A and B from the eye of B}), the variance of $\tilde{A}$ and $\tilde{B}$ are decomposed as 
\begin{eqnarray}
\sigma_{\tilde{A}}^2[\psi]&=&\Delta_{\tilde{A}}^2[\psi]+\mathcal{E}_{\tilde{A}}^2[\psi],\nonumber\\
\sigma_{\tilde{B}}^2[\psi]&=&\Delta_{\tilde{B}}^2[\psi]. 
\label{decomposition of the quantum variance of A and B into the variance of the c-valued quantities}
\end{eqnarray}

Next, since $\Delta_{\tilde{O}}^2[\psi]$ and $\mathcal{E}_{\tilde{O}}^2[\psi]$ are just the variances of c-valued | i.e., classical, commuting | variables, they must satisfy the usual uncertainty relation for real-valued classical random variables. To this end, recall that, from conventional probability theory, one has, using the Cauchy-Schwartz inequality, the following general relation between the variances of any two classical random variables $\mathcal{X}$ and $\mathcal{Y}$, and their corresponding covariance: 
\begin{eqnarray}
\braket{(\mathcal{X}-\braket{\mathcal{X}})^2}\braket{(\mathcal{Y}-\braket{\mathcal{Y}})^2}\ge|\braket{(\mathcal{X}-\braket{\mathcal{X}})(\mathcal{Y}-\braket{\mathcal{Y}})}|^2.
\label{general classical uncertainty relation between two c-numbers}
\end{eqnarray}
We show below that imposing the above classical uncertainty relation to the two c-valued physical quantities $\tilde{A}(b_n,\xi|\psi)$ and $\tilde{B}(b_n,\xi|\psi)$ in Eq. (\ref{A and B from the eye of B}) leads to the Schr\"odinger and Kennard-Robertson lower bounds in the KRS uncertainty relation, separately. 

First, taking $\mathcal{X}=A^{\rm w}_{\mathcal{R}}(b_n|\psi)$ and $\mathcal{Y}=B^{\rm w}_{\mathcal{R}}(b_n|\psi)=b_n$ into Eq. (\ref{general classical uncertainty relation between two c-numbers}), one has:
\begin{eqnarray}
&&\Delta_{\tilde{A}}^2[\psi]\Delta_{\tilde{B}}^2[\psi]\nonumber\\
&=&\braket{(A^{\rm w}_{\mathcal{R}}-\braket{A^{\rm w}_{\mathcal{R}}})^2}\braket{(B^{\rm w}_{\mathcal{R}}-\braket{B^{\rm w}_{\mathcal{R}}})^2}\nonumber\\
&\ge&\big|\braket{(A^{\rm w}_{\mathcal{R}}-\braket{A^{\rm w}_{\mathcal{R}}})(B^{\rm w}_{\mathcal{R}}-\braket{B^{\rm w}_{\mathcal{R}}})}\big|^2\nonumber\\
&=&\big|\braket{A^{\rm w}_{\mathcal{R}}B^{\rm w}_{\mathcal{R}}}-\braket{A^{\rm w}_{\mathcal{R}}}\braket{B^{\rm w}_{\mathcal{R}}}\big|^2\nonumber\\ 
&=&\Big|\sum_n{\rm Re}\Big\{\frac{\braket{b_n|\hat{A}|\psi}}{\braket{b_n|\psi}}\Big\}b_n\big|\braket{b_n|\psi}\big|^2-\braket{\psi|\hat{A}|\psi}\braket{\psi|\hat{B}|\psi}\Big|^2\nonumber\\
&=&\Big|\sum_n{\rm Re}\Big\{\braket{\psi|b_n|b_n}\braket{b_n|\hat{A}|\psi}\Big\}-\braket{\psi|\hat{A}|\psi}\braket{\psi|\hat{B}|\psi}\Big|^2\nonumber\\
&=&\big|\braket{\psi|\frac{1}{2}\{\hat{A},\hat{B}\}|\psi}-\braket{\psi|\hat{A}|\psi}\braket{\psi|\hat{B}|\psi}\big|^2,
\label{Schroedinger lower bound arising from weak values}
\end{eqnarray}
where we have used a corollary of Eq. (\ref{real part of quantum expectation value as mean value of real-global-stochastic variable}) that $\braket{O^{\rm w}_{\mathcal{R}}(b_n|\psi)}=\braket{\psi|\hat{O}|\psi}$. Next, putting $\mathcal{X}=(\xi/\hbar) A^{\rm w}_{\mathcal{I}}(b_n|\psi)$ and $\mathcal{Y}=B^{\rm w}_{\mathcal{R}}(b_n|\psi)=b_n$ into Eq. (\ref{general classical uncertainty relation between two c-numbers}), one obtains, 
\begin{eqnarray}
&&\mathcal{E}_{\tilde{A}}^2[\psi]\Delta_{\tilde{B}}^2[\psi]\nonumber\\
&=&\Big\langle\Big(\frac{\xi}{\hbar}A^{\rm w}_{\mathcal{I}}\Big)^2\Big\rangle\braket{(B^{\rm w}_{\mathcal{R}}-\braket{B^{\rm w}_{\mathcal{R}}})^2}\nonumber\\
&\ge&\big|\big\langle A^{\rm w}_{\mathcal{I}}(B^{\rm w}_{\mathcal{R}}-\braket{B^{\rm w}_{\mathcal{R}}})\big\rangle\big|^2\nonumber\\
&=&\Big|\sum_n{\rm Im}\Big\{\frac{\braket{b_n|\hat{A}|\psi}}{\braket{b_n|\psi}}\Big\}(b_n-\braket{\psi|\hat{B}|\psi})|\braket{b_n|\psi}|^2\Big|^2\nonumber\\
&=&\Big|\sum_n{\rm Im}\Big\{\braket{\psi|b_n\ket{b_n}\bra{b_n}\hat{A}|\psi}\Big\}\Big|^2\nonumber\\
&=&\frac{1}{4}\Big|\braket{\psi|[\hat{A},\hat{B}]|\psi}\Big|^2.
\label{Kennard-Robertson lower bound arising from the imaginary part of weak value}
\end{eqnarray}
Note importantly that in the above derivations, we have used the fact that both $\hat{A}$ and $\hat{B}$ are Hermitian. Hence, by taking the eigenstates $\{\ket{b_n}\}$ of $\hat{B}$ as the reference coordinate basis, it becomes clear that the Schr\"odinger lower bound arises from the fluctuation of the $\xi$-independent term, and the Kennard-Robertson lower bound arises from the fluctuation of the $\xi$-dependent error term of $\tilde{A}(b_n,\xi|\psi)$. We can instead of course take the complete set of eigenvectors $\{\ket{a}\}$ of $\hat{A}$ as the reference coordinate basis, so that instead of Eq. (\ref{A and B from the eye of B}) we obtain $\tilde{A}(a_n,\xi|\psi)=a_n$ and $\tilde{B}(a_n,\xi|\psi)=B^{\rm w}_{\mathcal{R}}(a_n|\psi)+\frac{\xi}{\hbar}B^{\rm w}_{\mathcal{I}}(a_n|\psi)$. In this way, we interchange the role of $\tilde{A}$ and $\tilde{B}$ in the above derivations leading to the same Schr\"odinger and Kennard-Robertson lower bounds. We emphasize that while all the terms on left-hand sides of Eqs. (\ref{Schroedinger lower bound arising from weak values}) and (\ref{Kennard-Robertson lower bound arising from the imaginary part of weak value}) depend on the choice of reference coordinate basis, hence on the choice of the post-selection measurement context to define the c-valued physical quantities via weak values, the right-hand sides are basis-independent. 

Using Eqs. (\ref{ER classical variance is equal to quantum variance}) and (\ref{decomposition of the quantum variance of A and B into the variance of the c-valued quantities}) for the decomposition of quantum variance into the fluctuations of the two decomposing terms of the c-valued physical quantities, and noting Eqs. (\ref{Schroedinger lower bound arising from weak values}) and (\ref{Kennard-Robertson lower bound arising from the imaginary part of weak value}), we thus finally obtain the KRS uncertainty relation
\begin{eqnarray}
&&\sigma_{\hat{A}}^2[\psi]\sigma_{\hat{B}}^2[\psi]=\sigma_{\tilde{A}}^2[\psi]\sigma_{\tilde{B}}^2[\psi]\nonumber\\
&=&\mathcal{E}_{\tilde{A}}^2[\psi]\Delta_{\tilde{B}}^2[\psi]+\Delta_{\tilde{A}}^2[\psi]\Delta_{\tilde{B}}^2[\psi]\nonumber\\
&\ge&\frac{1}{4}\big|\braket{\psi|[\hat{A},\hat{B}]|\psi}\big|^2\nonumber\\
&+&\Big|\frac{1}{2}\braket{\psi|\{\hat{A},\hat{B}\}|\psi}-\braket{\psi|\hat{A}|\psi}\braket{\psi|\hat{B}|\psi}\Big|^2. 
\label{Robertson-Schroedinger uncertainty relation from weak values}
\end{eqnarray}
Hence, the KRS uncertainty relation can be seen as the classical uncertainty relation imposed on to the c-valued physical quantities defined in Eq. (\ref{fundamental epistemic decomposition}) constructed from complex weak values with post-selection over a complete set of basis and a universal global random variable $\xi$. It thus reveals the irreducible statistical scatter of well defined real-deterministic quantities, rather than represents some irreducible random disturbance in measurement. Moreover, the result suggests a deep relation between the KRS uncertainty relation and the decomposition of the c-valued physical quantities into two distinct terms in Eq. (\ref{fundamental epistemic decomposition}) to be further elaborated in the next section. 

Within the representation based on c-valued physical quantities, it is clear that if there is a reference coordinate basis $\{\ket{\phi_n}\}$ which jointly diagonalizes $\hat{A}$ and $\hat{B}$, then the $\xi-$dependent error terms in both $\tilde{A}(\phi_n,\xi|\psi)$ and $\tilde{B}(\phi_n,\xi|\psi)$ defined in Eq. (\ref{fundamental epistemic decomposition}) vanish simultaneously. This implies that also $\mathcal{E}_{\tilde{A}}^2$ and $\mathcal{E}_{\tilde{B}}^2$ simultaneously vanish so that the Kennard-Robertson bound derived in Eq. (\ref{Kennard-Robertson lower bound arising from the imaginary part of weak value}) vanishes. Hence, the finite Kennard-Robertson lower bound arises because, for noncommuting observables $[\hat{A},\hat{B}]\neq 0$, there is no common reference coordinate basis $\{\ket{\phi_n}\}$ in which the error terms in both $\tilde{A}(\phi_n,\xi|\psi)$ and $\tilde{B}(\phi_n,\xi|\psi)$ simultaneously vanish. In this sense, the present representation thus captures the incompleteness of the Hilbert space basis to describe the quantum statistics via real-deterministic c-valued variables, and the need to introduce an auxiliary global random variable $\xi$. It is of course the characteristic of quantum mechanics that  there is no common basis wherein non-commuting observables are jointly determined (definite) via strong measurement. Note however that within the standard quantum mechanics, quantum randomness is commonly seen as objective. By contrast, within the representation based on the c-valued physical quantities developed in the present work, the randomness is due to the inaccessibility of $\xi$ by the agent. See also the discussion in Sec. \ref{Optimal estimation under epistemic restriction}.  

Let us mention that Hall {\it et al}. in Ref. \cite{Hall weak value to quantum uncertainty} has also derived the KRS uncertainty relation from the classical uncertainty relation imposed to deterministic classical variables. Their deterministic variables are however directly given by the complex weak value even for Hermitian observables. In this sense, they argue that the mystery of the quantum uncertainty relation lies in the complex valued-ness of the weak value. Consequently, however, it is not clear how the Kennard-Robertson and  Schr\"odinger lower bounds arise separately within the model, despite the fact that the two lower bounds are conceptually distinct. On the other hand, Lee {\it et al}. in Ref. \cite{Lee uncertainty relation from weak value} have derived separately the two lower bounds by using the weak value. However, their derivation fully employs the machinery of linear operators in Hilbert space so that its correspondence with classical uncertainty relation is not clear. By contrast, within our derivation, the Schr\"odinger and the Kennard-Robertson lower bounds arise separately respectively by imposing the classical uncertainty relation on to the $\xi$-independent term and the $\xi$-dependent error term of the real-deterministic c-valued physical quantities defined in Eq. (\ref{fundamental epistemic decomposition}).   

A different well-known method to map the q-valued quantum operators into c-valued variables is to employ the quasiprobability representations \cite{Hillery quasiprobability review,Lee quasiprobability review,Ferrie review quasiprobability}. In fact, quasiprobability representation is deliberately devised to express the quantum expectation values of arbitrary operators evocative of classical average over c-valued variables. This approach therefore also leads to the expression of the KRS uncertainty relation in terms of c-valued variables. However, the quasiprobability distributions associated with quantum preparation and measurement may take on negative or even complex values, or is highly irregular. Moreover, from classical statistical point of view, it is not fully clear how the uncertainty relation, and in particular the Schr\"odinger and Kennard-Robertson lower bounds, arise within the quasiprobability representations. We note that within our formalism, from Eq. (\ref{fundamental epistemic decomposition}),  for two arbitrary operators $\hat{A}$ and $\hat{B}$, a pre-selected state $\ket{\psi}$, and a reference coordinate basis $\{\ket{\phi_n}\}$, we always have a well-defined non-negative joint probability distribution of the associated c-valued physical quantities 
\begin{eqnarray}
&&{\rm Pr}(\tilde{A},\tilde{B}|\phi_n,\xi,\psi)\nonumber\\
&=&\delta\Big(\tilde{A}-\Big[A^{\rm w}_{\mathcal{R}}(\phi_n|\psi)+\frac{\xi}{\hbar}A^{\rm w}_{\mathcal{I}}(\phi_n|\psi)\Big]\Big)\nonumber\\
&\times&\delta\Big(\tilde{B}-\Big[B^{\rm w}_{\mathcal{R}}(\phi_n|\psi)+\frac{\xi}{\hbar}B^{\rm w}_{\mathcal{I}}(\phi_n|\psi)\Big]\Big). 
\label{joint distribution of c-valued physical quantities}
\end{eqnarray} 
We shall show below that when $\hat{A}$ and $\hat{B}$ are noncommuting, the above joint probability cannot take an arbitrary form, which, in turn, underlies the uncertainty relation. 

\section{Interpretation and discussion\label{Interpretation and discussion}}

\subsection{Quantum incompatibility as an irreducible epistemic restriction \label{Quantum incompatibility as epistemic restriction}}

Since we have derived the KRS uncertainty relation from classical uncertainty relation imposed on to the real-deterministic c-valued physical quantities defined in Eq. (\ref{fundamental epistemic decomposition}), we have thus shifted the nonclassicality associated with the quantum uncertainty relation to the structure of the c-valued physical quantities. Namely, it is not the relation which is classically inexplicable, but rather the c-valued physical quantity which is devised by recombining additively the real and imaginary parts of the weak value with the help of an auxiliary global random variable $\xi$. In particular, $\xi$ couples to the imaginary part of the weak value, transforming the incompatibility between the observable and the reference coordinate basis into the real-valued error term which is responsible for the Kennard-Robertson uncertainty relation. We shall argue below that the Kennard-Robertson uncertainty relation reflects a specific irreducible  interdependence between the c-valued physical quantities $\tilde{A}(\phi_n,\xi|\psi)$ and $\tilde{B}(\phi_n,\xi|\psi)$ when the associated Hermitian operators $\hat{A}$ and $\hat{B}$ are noncommuting. 

First, assume that the reference coordinate basis is given by the complete set of eigenvectors $\{\ket{b_n}\}$ of $\hat{B}$ with the associated set of eigenvalues $\{b_n\}$. Then, one has $\tilde{B}(b_n,\xi|\psi)=b_n$ which is distributed as ${\rm Pr}(b_n|\psi)=|\braket{b_n|\psi}|^2$. On the other hand, in this basis, the strength of the error term in the c-valued physical quantity $\tilde{A}(b_n,\xi|\psi)$ associated with a Hermitian operator $\hat{A}$, which is given by the imaginary part of the weak value $A^{\rm w}_{\mathcal{I}}(b_n|\psi)$, can be expressed as
\begin{eqnarray}
A^{\rm w}_{\mathcal{I}}(b_n|\psi)=\frac{\braket{b_n|[\hat{A},\ket{\psi}\bra{\psi}]|b_n}}{2i|\braket{b_n|\psi}|^2}=\frac{\hbar}{2}\frac{\frac{{\rm d}}{{\rm d}\theta}|\braket{b_n|\psi}|^2}{|\braket{b_n|\psi}|^2}. 
\label{imaginary term: variation of the reference coordinate basis which diagonalize B due to change of A}
\end{eqnarray}
Here, we have used $({\rm d}/{\rm d}\theta)\ket{\psi}\bra{\psi}\doteq-\frac{i}{\hbar}[\hat{A},\ket{\psi}\bra{\psi}]$ to arrive at the second equality, that is, we assumed that the Hermitian operator $\hat{A}$ generates a unitary transformation 
\begin{eqnarray}
\hat{U}=e^{-\frac{i}{\hbar}\hat{A}\theta},
\label{generator of unitary transformation}
\end{eqnarray}
to the  state $\ket{\psi}$ with the evolution parameter $\theta$. Equation (\ref{imaginary term: variation of the reference coordinate basis which diagonalize B due to change of A}) shows that the error term of the c-valued physical quantity $\tilde{A}(b_n,\xi|\psi)$ is proportional to the variation of the distribution of $\tilde{B}(b_n,\xi|\psi)=b_n$ with respect to the change of $\ket{\psi}$ induced by the unitary transformation of Eq. (\ref{generator of unitary transformation}). Upon inserting Eq. (\ref{imaginary term: variation of the reference coordinate basis which diagonalize B due to change of A}) into Eq. (\ref{fundamental epistemic decomposition}), the c-valued physical quantity $\tilde{A}(b_n,\xi|\psi)$ thus takes the form: 
\begin{eqnarray}
\tilde{A}(b_n,\xi|\psi)=A^{\rm w}_{\mathcal{R}}(b_n|\psi)+\frac{\xi}{2}\frac{\frac{{\rm d}}{{\rm d}\theta}|\braket{b_n|\psi}|^2}{|\braket{b_n|\psi}|^2}. 
\label{fundamental epistemic decomposition in terms of epistemic restriction}
\end{eqnarray}

Eq. (\ref{fundamental epistemic decomposition in terms of epistemic restriction}) reveals an intrinsic correlation or interdependence between the distribution of $\tilde{B}(b_n,\xi|\psi)=b_n$ which is given by ${\rm Pr}(b_n|\psi)=|\braket{b_n|\psi}|^2$, and the functional form of $\tilde{A}(b_n,\xi|\psi)$. This is a form of epistemic or statistical restriction, in the sense that the allowed distribution of $\tilde{B}(b_n,\xi|\psi)=b_n$ is restricted or parameterized by the functional form of $\tilde{A}(b_n,\xi|\psi)$. This in turn will therefore restrict the allowed joint distribution of $(\tilde{A},\tilde{B})$ defined in Eq. (\ref{joint distribution of c-valued physical quantities}). Note that when this epistemic restriction vanishes, namely when $A^{\rm w}_{\mathcal{I}}(b_n|\psi)=0$ for all $b_n$, one has $[\hat{A},\hat{\Pi}_{b_n}]=0$ for all $b_n$ so that $\hat{A}$ and $\hat{B}$ are commuting: $[\hat{A},\hat{B}]=\sum_nb_n[\hat{A},\hat{\Pi}_{b_n}]=0$. From this observation, the Kennard-Robertson uncertainty relation derived in the previous section can thus be seen to reflect such an irreducible epistemic restriction. 

To get a better intuition on such a specific epistemic restriction, let us compare it with what happens in classical mechanics. For this purpose, take the complete set of eigenvectors $\{\ket{q}\}$ of the position operator $\hat{q}$ as the coordinate basis, so that we have $\tilde{q}(q,\xi|\psi)=q$. Now, let us again write the wave function associated with the pre-selected state $\ket{\psi}$ in polar form, i.e., $\psi(q)=\braket{q|\psi}=\sqrt{\rho(q)}e^{iS(q)/\hbar}$ so that we have ${\rm Pr}(\tilde{q}|\psi)=|\psi(q)|^2=\rho(q)$. The c-valued momentum in the position coordinate basis then reads
\begin{eqnarray}
\tilde{p}(q,\xi|\psi)=\partial_qS(q)-\frac{\xi}{2}\frac{\partial_q\rho(q)}{\rho(q)}.
\label{fundamental epistemic decomposition position-momentum} 
\end{eqnarray}
Note that the unitary operator $\hat{U}=e^{-\frac{i}{\hbar}\hat{p}q_o}$ applied to the pre-selected state $\ket{\psi}$ generates a spatial shift $q_o$ to the wave function $\psi(q)$, so that we have $\partial_q\rho=-\partial_{q_o}\rho$. Upon inserting into Eq. (\ref{fundamental epistemic decomposition position-momentum}), we obtain $\tilde{p}(q,\xi|\psi)=\partial_qS(q)+\frac{\xi}{2}\frac{\partial_{q_o}\rho(q)}{\rho(q)}$, in accord with the general relation of Eq. (\ref{fundamental epistemic decomposition in terms of epistemic restriction}). Noting this, the error term of the c-valued momentum, i.e., the second term on the right-hand side of Eq. (\ref{fundamental epistemic decomposition position-momentum}), thus reflects the variation of the $\rho(q)$ due to the spatial shift of the wave function induced by the unitary transformation $\hat{U}=e^{-\frac{i}{\hbar}\hat{p}q_o}$ generated by $\hat{p}$. 

One can then see from Eq. (\ref{fundamental epistemic decomposition position-momentum}) that the allowed distribution of position $\rho(q)$ is indeed inherently restricted or fundamentally parameterized by the form of the underlying momentum field $\tilde{p}(q,\xi|\psi)$. Operationally, it means that each spatial point $q$ cannot be sampled with an arbitrary weight $\rho(q)$ independent of the momentum field $\tilde{p}(q,\xi|\psi)$ \cite{Agung-Daniel model}. In a sharp contrast to this, in classical mechanics, the distribution of position can always be chosen freely independent of the underlying momentum field, i.e., each spatial point $q$ in the momentum field can be weighted with an arbitrary $\rho(q)$. In this sense, classical mechanics thus enjoys an epistemic freedom which is violated by a specific epistemic restriction embodied in Eq. (\ref{fundamental epistemic decomposition position-momentum}). This epistemic restriction in turn constrains the allowed form of the joint distribution of $(\tilde{q},\tilde{p})$. 

As an illustrative example of such a statistical constraint in phase space, assume that the c-valued momentum takes the form $\tilde{p}(q,\xi|\psi)=\partial_qS=p_o$, where $p_o$ is a constant. Hence, the $\xi$-dependent term is vanishing and the c-valued momentum has a distribution with a sharp peak at $\tilde{p}=p_o$. Then, to comply with Eq. (\ref{fundamental epistemic decomposition position-momentum}), we must have $\partial_q\rho/\rho=0$ so that $\rho(q)$ must be spatially uniform. Moreover, one also has $S(q)=p_oq$, up to an unimportant constant. The above situation exactly captures the case when the wave function is a plane wave, i.e., $\psi(q)\sim e^{ip_oq/\hbar}$. Hence, when the distribution of $\tilde{p}$ is sharp, $\rho(\tilde{q})=\rho(q)$ must be spatially uniform in accordance to the spirit of the Heisenberg uncertainty principle. Indeed, from the specific interdependence between the momentum field and the distribution of position expressed in Eq. (\ref{fundamental epistemic decomposition position-momentum}), one can directly derive the KRS uncertainty relation for position and momentum \cite{Agung-Daniel model,Agung estimation independence,Agung nonlinear Schrodinger equation and EI}: 
\begin{eqnarray}
&&\sigma_{\tilde{q}}^2[\psi]\sigma_{\tilde{p}}^2[\psi]\nonumber\\
&\ge&\big|\frac{1}{2}\braket{\psi|\{\hat{q},\hat{p}\}|\psi}-\braket{\psi|\hat{q}|\psi}\braket{\psi|\hat{p}|\psi}\big|^2+\frac{\hbar^2}{4},
\label{KRS uncertainty relation for position and momentum}
\end{eqnarray} 
where we have made use of Eq. (\ref{Planck constant}). 

\subsection{Optimal estimation under epistemic restriction, Kennard-Robertson uncertainty relation, and classical limit \label{Optimal estimation under epistemic restriction}}

With the above observation in mind, below we shall argue that the two terms decomposing the c-valued physical quantity in Eq. (\ref{fundamental epistemic decomposition}) can be interpreted as an epistemic decomposition describing an optimal (i.e., best) estimation under the epistemic restriction. Let us take again the complete set of eigenvectors $\{\ket{b_n}\}$ of $\hat{B}$ as the reference coordinate basis so that we have $\tilde{B}(b_n,\xi|\psi)=b_n$. Now, consider an agent who wishes to estimate $\tilde{A}(b_n,\xi|\psi)$ associated with a Hermitian operator $\hat{A}$ and a pre-selected state $\ket{\psi}$, based on the information about $b_n$ sampled from the distribution ${\rm Pr}(b_n|\psi)=|\braket{b_n|\psi}|^2$, which, as argued in Sec. \ref{Quantum incompatibility as epistemic restriction}, is fundamentally parameterized by $\tilde{A}(b_n,\xi|\psi)$. Assume that the agent's estimate is given by 
\begin{eqnarray}
A^{\rm w}_{\mathcal{R}}(b_n|\psi), \nonumber
\end{eqnarray}
i.e., the first term on the right-hand side of Eq. (\ref{fundamental epistemic decomposition}). The single-shot estimation error is thus given by the second term on the right-hand side of Eq. (\ref{fundamental epistemic decomposition}), i.e., 
\begin{eqnarray}
\epsilon_{\tilde{A}}(b_n,\xi|\psi)&\doteq&\tilde{A}(b_n,\xi|\psi)-A^{\rm w}_{\mathcal{R}}(b_n|\psi)\nonumber\\
&=&\frac{\xi}{\hbar}A^{\rm w}_{\mathcal{I}}(b_n|\psi)=\frac{\xi}{\hbar}\frac{\braket{\psi|[\hat{\Pi}_{b_n},\hat{A}]|\psi}}{2i|\braket{b_n|\psi}|^2}. 
\label{single-shot estimation error}
\end{eqnarray} 
For any $\xi$, the average of the error over the distribution of $b_n$ is vanishing: $\sum_n\epsilon_{\tilde{A}}(b_n,\xi|\psi)|\braket{b_n|\psi}|^2=\frac{\xi}{2i\hbar}\sum_n\braket{\psi|[\hat{\Pi}_{b_n},\hat{A}]|\psi}=0$. In this sense, the above estimation scheme is unbiased. 

One can then see from Eq. (\ref{single-shot estimation error}) that the estimation error is larger when the incompatibility between $\hat{A}$ and the reference coordinate basis $\{\ket{b_n}\}$ over the state $\ket{\psi}$ is stronger. Computing the mean squared (MS) estimation error one gets
\begin{eqnarray}
\sum_n\int{\rm d}\xi \epsilon_{\tilde{A}}(b_n,\xi|\psi)^2\chi(\xi)|\braket{b_n|\psi}|^2\nonumber\\
=\big\langle A^{\rm w}_{\mathcal{I}}(b_n|\psi)^2\big\rangle=\mathcal{E}_{\tilde{A}}^2[\psi], 
\label{MS estimation error of A within the reference of B}
\end{eqnarray} 
where we have made use of Eq. (\ref{precision and accuracy}) in the last equality. Within the above interpretation, the decomposition of quantum variance in Eq. (\ref{decomposition of the quantum variance into the variance of the c-valued quantities}) can therefore be read as the decomposition of the variance of the c-valued physical quantity into the precision (i.e., the spread) of the estimate given by $\Delta_{\tilde{A}}^2[\psi]$ and the accuracy of estimation given by the associated MS estimation error $\mathcal{E}_{\tilde{A}}^2[\psi]$. 

Let us show that the above estimation scheme is optimal, i.e., it minimizes the MS estimation error. First, given a pre-selected state $\ket{\psi}$ and a reference coordinate basis $\{\ket{b_n}\}$, let $T_{\tilde{A}}(b_n|\psi)$ be an estimator for $\tilde{A}(b_n,\xi|\psi)$ based on information on $b_n$. The associated MS estimation error can be directly computed to give, noting Eqs. (\ref{fundamental epistemic decomposition}) and (\ref{Planck constant}), 
\begin{eqnarray}
&&\big\langle(\tilde{A}(b_n,\xi|\psi)-T_{\tilde{A}}(b_n|\psi))^2\big\rangle\nonumber\\
&=&\Big\langle\Big(A^{\rm w}_{\mathcal{R}}(b_n|\psi)-T_{\tilde{A}}(n|\psi)+\frac{\xi}{\hbar}A^{\rm w}_{\mathcal{I}}(b_n|\psi)\Big)^2\Big\rangle\nonumber\\
&=&\big\langle(T_{\tilde{A}}(n|\psi)-A^{\rm w}_{\mathcal{R}}(b_n|\psi))^2\big\rangle+\big\langle(A^{\rm w}_{\mathcal{I}}(b_n|\psi))^2\big\rangle. 
\label{proof of optimal-best estimate}
\end{eqnarray}    
One can see that, as claimed, the minimum is obtained when the estimator is given by 
\begin{eqnarray}
T_{\tilde{A}}(b_n|\psi)=A^{\rm w}_{\mathcal{R}}(b_n|\psi),
\end{eqnarray}
so that the first term on the right-hand side of Eq. (\ref{proof of optimal-best estimate}) vanishes. Moreover, the minimum MS estimation error is indeed given by Eq. (\ref{MS estimation error of A within the reference of B}). 

This interpretation of the real and imaginary parts of the weak value which constitute the c-valued physical quantity defined in Eq. (\ref{fundamental epistemic decomposition}) is similar to that proposed by Johansen \cite{Johansen weak value best estimation} and Hall \cite{Hall exact UR,Hall prior information}. See also Refs. \cite{Lee uncertainty relation from weak value,Dressel weak value as quantum interference,Hofmann contextual value - weak value}. Note however that, there, the authors consider the estimation of the observable $\hat{A}$ based on the measurement of another observable $\hat{B}$, whereas, here, we make an estimate of the c-valued physical quantity $\tilde{A}(b_n,\xi|\psi)$ based on information on another c-valued physical quantity $\tilde{B}(b_n,\xi|\psi)=b_n$. Accordingly, the definition and evaluation of the MS estimation error is different: while in our approach it is simply defined as the $l_2$-norm between the c-valued physical quantities with well-defined operational meaning, in Refs. \cite{Johansen weak value best estimation,Hall exact UR,Hall prior information}, it is defined based on the statistical deviation between Hermitian operators over a quantum state. While both definitions are formally shown to be equivalent via Eq. (\ref{l_2 norm between two c-valued physical quantities is equal to the statistical deviation between operators}), the intuition and interpretation that come with them are clearly different. Moreover, while the authors in Refs. \cite{Johansen weak value best estimation,Hall exact UR,Hall prior information} identify the imaginary part of the weak value as the square root of the MS error of the estimation, they did not provide the single-shot estimation error. Within our formalism, the proposed single-shot estimation error of Eq. (\ref{single-shot estimation error}) is made possible via the introduction of a global random variable $\xi$. We expect that such a supplementary variable may lead to predictions beyond quantum mechanics.  

The above results suggest that the decomposition of the c-valued physical quantity into two terms in Eq. (\ref{fundamental epistemic decomposition}) is epistemic (i.e., informational) not physical. Namely, the two terms on the right-hand side of Eq. (\ref{fundamental epistemic decomposition}) are not agent-independent real-physical quantities describing a random physical noise (the second term) superimposed onto a deterministic physical fluctuation (the first term). Rather, they are artificially devised by an agent to optimally infer the value of a physical quantity compatible with her information obtained in preparation (pre-selection) and measurement (post-selection). This is the reason why in Sec. \ref{c-valued physical quantities from weak values and an auxiliary global random variable} we called the $\xi$-dependent term in Eq. (\ref{fundamental epistemic decomposition}) as an epistemic ``error term'' rather than as a physical ``noise term.''  Recall that the first term and the magnitude of the second term of the right-hand side of Eq. (\ref{fundamental epistemic decomposition}) can be obtained separately via weak measurement from the average shift of position and momentum of the pointer. This suggests that they are (i.e., the estimate and the estimation error) somehow uncorrelated, which is a desirable feature from the point of view of information theory. As an example, the c-valued momentum can be decomposed as in Eq. (\ref{fundamental epistemic decomposition position-momentum}) where the first term is the optimal estimate of $\tilde{p}$ given $q$ and the second term is the single-shot estimation error. One can see that the two terms are independent of each other, wherein $S(q)$ is the phase and $\rho(q)$ is the amplitude of the wave function $\psi(q)=\sqrt{\rho(q)}e^{iS(q)/\hbar}$ associated with the preparation. 

Next, let us again recall that within the basis $\{\ket{b_n}\}$, the c-valued physical quantity associated with $\hat{B}$ takes the form $\tilde{B}(b_n,\xi|\psi)=B^{\rm w}_{\mathcal{R}}(b_n|\psi)=b_n$. By definition, it can be seen as the unbiased estimate of its own mean value, i.e.,  $B_o[\psi]\doteq\braket{\tilde{B}(b_n,\xi|\psi)}=\braket{B^{\rm w}_{\mathcal{R}}(b_n|\psi)}$, parameterizing the distribution of $\tilde{B}(b_n,\xi|\psi)=b_n$ given by ${\rm Pr}(b_n|\psi)=|\braket{b_n|\psi}|^2$. The associated MS error of the estimation thus reads  
\begin{eqnarray}
\mathcal{E}_{B_o}^2[\psi]&\doteq&\big\langle(\tilde{B}(b_n,\xi|\psi)-B_o[\psi])^2\big\rangle\nonumber\\
&=&\Big\langle\big(B^{\rm w}_{\mathcal{R}}(b_n|\psi)-\braket{B^{\rm w}_{\mathcal{R}}(b_n|\psi)}\big)^2\Big\rangle\nonumber\\
&=&\Delta_{\tilde{B}}^2[\psi], 
\label{estimation error of B within the reference of B}
\end{eqnarray}  
where the last equality is just Eq. (\ref{precision and accuracy}). Multiplying $\mathcal{E}_{\tilde{A}}^2[\psi]$ to both sides of Eq. (\ref{estimation error of B within the reference of B}) and noting Eq. (\ref{Kennard-Robertson lower bound arising from the imaginary part of weak value}), we thus obtain
\begin{eqnarray}
\mathcal{E}_{\tilde{A}}^2[\psi]\mathcal{E}_{B_o}^2[\psi]=\mathcal{E}_{\tilde{A}}^2\Delta_{\tilde{B}}^2[\psi]\ge\frac{1}{4}\big|\braket{\psi|[\hat{A},\hat{B}]|\psi}\big|^2.
\label{Kennard-Robertson uncertainty as trade-off between MS estimation errors of joint estimations}
\end{eqnarray}
In this sense, by interpreting $\tilde{A}(b_n,\xi|\psi)$ and $B_o[\psi]$ as the parameters for the distribution ${\rm Pr}(b_n|\psi)=|\braket{b_n|\psi}|^2$, the Kennard-Robertson lower bound can be seen as a fundamental limitation of the simultaneous estimation of the two parameters based on information on $\tilde{B}(b_n,\xi|\psi)=b_n$. In particular, it describes the trade-off between the associated MS estimation errors.  

Finally, within the above epistemic interpretation of the decomposition of the c-valued physical quantity in terms of the estimation under epistemic restriction, the limit of vanishing estimation error is obtained when $\xi\rightarrow 0$. In this limit, information on the coordinate basis $\{\ket{\phi_n}\}$ is sufficient to determine the c-valued physical quantities associated with arbitrary operators as in classical mechanics. Of particular interest is the estimation of the c-valued momentum $\tilde{p}(q,\xi|\psi)$ based on information on the conjugate position $\tilde{q}(q,\xi|\psi)=q$. In this case, the c-valued momentum is decomposed as in Eq. (\ref{fundamental epistemic decomposition position-momentum}), wherein the first term on the right-hand side is the optimal estimate and the second term is the single-shot estimation error. In the limit $\xi\rightarrow 0$, which is equivalent to $\hbar\rightarrow 0$ as per Eq. (\ref{Planck constant}), the estimation error is vanishing so that we obtain 
\begin{eqnarray}
\lim_{\xi\rightarrow 0}\tilde{p}(q,\xi|\psi)=\partial_qS(q). 
\label{Hamilton-Jacobi relation}
\end{eqnarray}
Hence, in this physical regime, the epistemic restriction disappears, namely, the momentum field no longer restricts the allowed form of the distribution of position as in classical mechanics. 

Moreover, upon identifying $S(q)$ as the Hamilton's principal function, Eq. (\ref{Hamilton-Jacobi relation}) is just the fundamental formula for the momentum field in classical mechanics. Noting this, the second term in Eq. (\ref{fundamental epistemic decomposition position-momentum}), i.e., the error of estimation of momentum given information on the conjugate position, which is also responsible for the epistemic restriction, can thus be thought of as a microscopic deviation from the classical formula of Eq. (\ref{Hamilton-Jacobi relation}). Indeed, in Ref. \cite{Agung-Daniel model}, we have shown that promoting the modified classical relation of Eq. (\ref{fundamental epistemic decomposition position-momentum}) as a fundamental axiom reflecting an irreducible epistemic restriction, and imposing the principles of conservation of average energy and trajectories, single out the Schr\"odinger equation for bosonic systems, which reduces to the classical Hamilton-Jacobi equation in the limit $\xi\rightarrow 0$. This mathematical result has recently motivated one of us to develop an epistemic interpretation of the abstract rules of quantum mechanics as a calculus for describing an estimation of momentum given information on the conjugate position under an irreducible epistemic restriction, satisfying an inferential principle of estimation independence \cite{Agung epistemic interpretation,Agung estimation independence,Agung nonlinear Schrodinger equation and EI}. 

\section{Summary and Future outlook \label{conclusion}}

Can we understand quantum uncertainty relations in terms of the fluctuations of real-deterministic c-valued (i.e., classical, commuting) variables? What fundamental features of classical mechanics do we need to give in for real-deterministic variables to respect the Heisenberg uncertainty principle?  Better understanding of these conceptual questions is pivotal to have a deeper insight into the nature of randomness in quantum world relative to that in classical world, and specifically how quantum uncertainty relation differs fundamentally from the uncertainty relation for classical random variables. In turn, it may also provide useful intuition to harness quantum uncertainty for practical applications in information processings. In the present work, given a pre-selected state $\ket{\psi}$, a general quantum operator $\hat{O}$ and a complete set of Hilbert space vectors $\{\ket{\phi_n}\}$ as a reference coordinate basis, we devised a class of real-deterministic c-valued physical quantities $\tilde{O}(\phi_n,\xi|\psi)$ which always satisfy the Kennard-Robertson-Schr\"odinger (KRS) uncertainty relation. Remarkably, this real-deterministic variable can be constructed from the complex weak value obtained operationally via weak measurement of $\hat{O}$ without significantly perturbing the pre-selected state $\ket{\psi}$ and followed by a post-selection over a complete set of state vectors $\{\ket{\phi_n}\}$. This is done by introducing an auxiliary global random variable $\xi$ to separate the real and imaginary parts of the complex weak value. 

The global random variable $\xi$ couples to the imaginary part of the weak value, transforming the incompatibility between the quantum operator and the reference coordinate basis into the strength of the fluctuations of a random error term. This $\xi$-dependent term is then superimposed onto a $\xi$-independent term given by the real-part of the weak value. Within the representation, the complementarity between two incompatible quantum observables is reflected in the trade-off between the strength of the $\xi$-dependent error terms of the associated two c-valued physical quantities. Hence, the quantum uncertainty relation between two incompatible observables arises because there is no common reference coordinate basis $\{\ket{\phi_n}\}$ so that the error terms of the associated two c-valued physical quantities vanish simultaneously. Furthermore, we have argued that the incompatibility between two quantum observables manifests a specific irreducible epistemic restriction between the associated two c-valued physical quantities, constraining the allowed form of their joint distribution, underlying the uncertainty relation. The KRS uncertainty relation thus describes a fundamental epistemic limitation on any attempt to define real-deterministic c-valued physical quantities to describe quantum statistics, i.e., that such c-valued physical quantities must satisfy a specific epistemic restriction, violating the epistemic freedom underlying the classical mechanics. 

We have shown that the class of c-valued physical quantities provide a real-deterministic contextual hidden variable model for the quantum expectation value of a certain class of operators. The contexts are determined by the choice of reference basis for post-selection measurement, the choice of weak measurement when the observable is a product of two observables, and the choice of preparation associated with a mixed state. The results suggest to promote the c-valued physical quantity $\tilde{O}(\phi_n,\xi|\psi)$ as a fundamental physical concept independent of any measurement model, i.e., as the counterfactual physical quantity complying with the boundary conditions given by the preparation and the choice of coordinate basis. A similar speculation concerning the weak value is suggested in Ref. \cite{Hosoya-Shikano counterfactual value} and is applied to discuss the Hardy paradox. Note however that, as argued in Sec. \ref{Quantum incompatibility as epistemic restriction}, even for such a realistic interpretation of $\tilde{O}(\phi_n,\xi|\psi)$, its decomposition into two parts, i.e., into the $\xi$-independent and $\xi$-dependent terms in Eq. (\ref{fundamental epistemic decomposition}), is epistemic (informational). Namely, the $\xi$-independent term should be interpreted as the optimal estimate of the c-valued physical quantity, while the $\xi$-dependent term is the associated single-shot estimation error. Hence, the specific decomposition in Eq. (\ref{fundamental epistemic decomposition}) is artificially made in the agent's mind to optimally organize her experience. Within this epistemic interpretation, as shown in Eq. (\ref{Kennard-Robertson uncertainty as trade-off between MS estimation errors of joint estimations}), the Kennard-Robertson uncertainty relation describes the trade-off between the MS errors of the joint estimation of two parameters associated with two noncommuting observables.  

Since the definition of c-valued quantities depends on the context determined by the choice of the reference coordinate basis for post-selection measurement, it is not sufficient for a measurement independent reformulation of quantum mechanics. However, c-valued quantities may be defined as counterfactual entities independent of measurement by choosing a physically meaningful reference basis. It may then be possible to reconstruct the whole of quantum mechanics, not just the quantum uncertainty relations, from scratch, starting from the c-valued physical quantities and imposing a set of plausible physical principles. For infinite dimensional continuous variable quantum systems, this program was shown to be viable in Ref. \cite{Agung-Daniel model}. In this reconstruction, we have assumed the c-valued momentum $\tilde{p}(q,\xi|\psi)$ given in Eq. (\ref{fundamental epistemic decomposition position-momentum}) as the counterfactual momentum field associated with a preparation. It is interesting to ask if this reconstruction program can be generalized to discrete variable quantum systems with finite dimension of Hilbert space by employing the c-valued physical quantity defined in Eq. (\ref{fundamental epistemic decomposition}), and if the specific forms of the two decomposing terms in Eq. (\ref{fundamental epistemic decomposition}) can be justified by the principle of estimation independence as for the two terms in Eq. (\ref{fundamental epistemic decomposition position-momentum}) \cite{Agung estimation independence,Agung nonlinear Schrodinger equation and EI}. Such a reconstruction will lend further support to the epistemic interpretation of quantum mechanics advanced in Ref. \cite{Agung epistemic interpretation}. Moreover, it is intriguing to further study if the counterfactual c-valued physical quantities can shed some fresh light in the discussion of the Hardy paradox \cite{Hardy paradox} relative to that based directly on the weak value \cite{Aharonov weak value Hardy paradox,Lundeen weak value Hardy paradox,Hosoya-Shikano counterfactual value,Yokota weak value Hardy paradox,Molmer negative population}. 

Several other questions are worth pursuing in the future. First, by introducing a supplementary global random variable $\xi$, we have been able to transform the complementarity between two incompatible quantum observables into the complementarity between the error terms of the associated two c-valued physical quantities. It is thus tempting to exploit this intuition to study the wave-particle duality and quantum coherence. It is also intriguing to explore the possible empirical implications of such a global random variable. Especially, assuming that $\xi$ characterizes the randomness of each single repetition of the weak measurement, we want to see if $\tilde{O}(\phi_n,\xi|\psi)$ is realizable in a single-shot weak measurement with post-selection and to study its distribution. We also would like to make a detailed comparison between the representation based on c-valued physical quantities with the approach based on quasiprobability, and to elaborate its computational possibility \cite{Agung efficient computation with ER ensemble}. In particular it is interesting the compare the present formalism with the complex-valued Kirkwood-Dirac quasiprobability \cite{Kirkwood KD quasiprobability distribution,Dirac KD quasiprobability distribution,Barut KD quasiprobability distribution}, or its real part, i.e., the Terletsky-Margenou-Hill quasiprobability \cite{Terletsky TMH quasiprobability distribution,Margenau-Hill TMH quasiprobability distribution}, which are also deeply related to the concept of weak value \cite{Johansen KD quasiprobability distribution,Hofmann KD quasiprobability distribution,Dressel weak value as quantum interference,Steinberg tunneling time via weak value 2,Steinberg tunneling time via weak value 1}. Finally, we want to further investigate the interpretation of the c-valued physical quantity in terms of estimation under epistemic restriction and its possible application in quantum parameter estimation \cite{Paris quantum estimation review,Giovannetti quantum estimation review}.

\begin{acknowledgments}   

\end{acknowledgments}   

\appendix 

\section{Derivation of Eq. (\ref{real-global-stochastic representation of quantum correlation}) \label{C-valued representation of quantum expectation value of product of two operators}}

Let us assume additionally that the third moment of $\xi$ is vanishing, i.e., $\overline{\xi^3}=0$. For example, assume that $\xi$ is a binary random variable $\xi=\pm\hbar$ with equal probability. We then obtain the following result concerning the representation of the commutator between two operators in terms of the statistics of the associated real-deterministic c-valued physical quantities:  
\begin{eqnarray}
&&\Big\langle\frac{\xi}{\hbar}\tilde{A}(\phi_n,-\xi|\psi)\tilde{B}(\phi_n,\xi|\psi)\Big\rangle\nonumber\\
&=&\sum_n\big(A^{\rm w}_{\mathcal{R}}(\phi_n|\psi)B^{\rm w}_{\mathcal{I}}(\phi_n|\psi)-A^{\rm w}_{\mathcal{I}}(\phi_n|\psi)B^{\rm w}_{\mathcal{R}}(\phi_n|\psi)\big)|\braket{\phi_n|\psi}|^2\nonumber\\
&=&\sum_n{\rm Im}\Big\{\frac{\braket{\phi_n|\hat{A}|\psi}^*}{\braket{\phi_n|\psi}^*}\frac{\braket{\phi_n|\hat{B}|\psi}}{\braket{\phi_n|\psi}}\Big\}|\braket{\phi_n|\psi}|^2\nonumber\\
&=&\frac{1}{2i}\braket{\psi|(\hat{A}^{\dagger}\hat{B}-\hat{B}^{\dagger}\hat{A})|\psi},
\label{real-global-stochastic representation of quantum commutator: general operators}
\end{eqnarray}
where to obtain the second equality we have used the mathematical identity ${\rm Im}\{a^*b\}={\rm Re}\{a\}{\rm Im}\{b\}-{\rm Im}\{a\}{\rm Re}\{b\}$ for two complex numbers $a$ and $b$. As a corollary, if both $\hat{A}$ and $\hat{B}$ are Hermitian, one thus gets
\begin{eqnarray}
\Big\langle\frac{\xi}{\hbar}\tilde{A}(\phi_n,-\xi|\psi)\tilde{B}(\phi_n,\xi|\psi)\Big\rangle=\frac{1}{2i}\braket{\psi|[\hat{A},\hat{B}]|\psi}. 
\label{real-global-stochastic representation of quantum commutator: Hermitian operators}
\end{eqnarray}
If furthermore the two Hermitian operators are commuting, we obtain $\braket{\frac{\xi}{\hbar}\tilde{A}(\phi_n,-\xi|\psi)\tilde{B}(\phi_n,\xi|\psi)}=0$. 

From Eqs. (\ref{real-global-stochastic representation of quantum anticommutator: general operators}) and (\ref{real-global-stochastic representation of quantum commutator: general operators}), we can thus compute the quantum expectation value of the product of two quantum operators as 
\begin{eqnarray}
\braket{\psi|\hat{A}^{\dagger}\hat{B}|\psi}&=&\braket{\tilde{A}(\phi_n,\xi|\psi)\tilde{B}(\phi_n,\xi|\psi)}\nonumber\\
&+&i\Big\langle\frac{\xi}{\hbar}\tilde{A}(\phi_n,-\xi|\psi)\tilde{B}(\phi_n,\xi|\psi)\Big\rangle, \nonumber
\label{real-global-stochastic representation of quantum correlation appendix}
\end{eqnarray}
as claimed in Eq. (\ref{real-global-stochastic representation of quantum correlation}). We note that the decomposition on the right-hand side is basis-dependent, i.e., contextual with respect to the choice of post-selection measurement $\{\ket{\phi_n}\}$ of the associated weak value measurement, while the left-hand side is not. One can get a simplified expression by taking the complete eigenstates $\{\ket{a_n}\}$ of $\hat{A}$ as the reference coordinate basis. The right-hand side of Eq. (\ref{real-global-stochastic representation of quantum correlation}) can thus be seen as a correlation between $\tilde{A}(a_n,\xi|\psi)=a_n$ and $\tilde{B}(a_n,\xi|\psi)+i\frac{\xi}{\hbar}\tilde{B}(b_n,\xi|\psi)$.

\end{document}